# Effects of Static and Conversational AI-Generated Messages on Colorectal Cancer Screening Intent: a Randomized Clinical Trial

Neil K. R. Sehgal, ME[1,2], Manuel Tonneau, MSc[3,4,5], Andy SL Tan[6], Shivan J. Mehta[2,7,8], Alison Buttenheim, PhD[2,9,10], Lyle Ungar, PhD[1], Anish K. Agarwal MD, MPH, MS[2,7,11], Sharath Chandra Guntuku, PhD[1,2]

[1]Computer & Information Science Department, University of Pennsylvania, Philadelphia, PA, USA
[2]Leonard Davis Institute of Health Economics, University of Pennsylvania, Philadelphia, PA, USA
[3]Oxford Internet Institute, University of Oxford, Oxford, United Kingdom
[4]Development Impact Department, World Bank, Washington, DC, USA
[5]Department of Computer Science, New York University, New York, NY, USA
[6]Annenberg School of Communication, University of Pennsylvania, Philadelphia, PA, USA
[7]Penn Medicine Center for Health Care Transformation and Innovation, University of Pennsylvania, Philadelphia, PA, USA
[8]Department of Medicine, University of Pennsylvania Perelman School of Medicine, Philadelphia, PA USA
[9]Department of Family and Community Health, School of Nursing, University of Pennsylvania, Philadelphia, PA, USA
[10]Center for Health Incentives and Behavioral Economics, University of Pennsylvania, Philadelphia, PA,
[11]Department of Emergency Medicine, University of Pennsylvania Perelman School of Medicine, Philadelphia, PA, US

**ABSTRACT**

Large language model chatbots show increasing promise in persuasive communication, but their clinical utility remains uncertain, particularly in settings where sustained conversations are difficult to scale. In a randomized clinical trial, 915 U.S. adults (ages 45-75) who had never completed colorectal cancer (CRC) screening were randomized to no-message control, expert-written patient materials, single AI-generated message, or motivational interviewing (MI) chatbot. Both AI interventions significantly increased stool-test intentions by over 12 points (12.9-13.8/100), compared to a 7.5 gain for expert materials ($p<.001$). While AI interventions outperformed the no-message control for colonoscopy intent, neither showed improvement over expert materials. Notably, we did not detect a statistically significant difference between the longer, interactive chatbot and the simpler single AI message, despite participants spending 3 minutes more on average engaging with it. These findings suggest concise, demographically tailored AI messages may offer a scalable approach to increasing CRC screening intentions, although effects on completed screening remain unknown.

**Trial Registration**
ClinicalTrials.gov Identifier: NCT07107152

**Main Text**

## INTRODUCTION

Colorectal cancer (CRC) remains a leading cause of cancer mortality in the United States, with over 150,000 new cases and over 52,000 deaths projected for 2025.[1] In 2021 the US Preventive Services Task Force (USPSTF) lowered the recommended starting age from 50 to 45 years and reiterated that multiple modalities (FIT, stool DNA-FIT, colonoscopy, flexible sigmoidoscopy, CT colonography) are equally acceptable for average-risk patients.[2] Despite the clear benefits of screening, nearly one-third of adults aged 50–75—and an even larger share of the newly eligible 45- to 49-year group—are not up-to-date.[3-4] As clinicians and health systems accelerate outreach beyond traditional clinic visits, and new testing modalities become accepted and available, patients must navigate the variety of choices, underscoring the need for scalable, comprehensible, and empathetic patient-centered tools.

Large language models (LLMs) have demonstrated growing persuasive power outside of healthcare: in a preregistered online debate platform, GPT-4 significantly increased the odds of attitude change relative to human opponents when allowed basic demographic targeting.[5] Within health care, however, a more nuanced picture continues to emerge. A single GPT-3–generated social-media post outperformed Centers for Disease Control and Prevention (CDC) posts in promoting vaccination, yet a three-minute HPV-vaccine chatbot dialog failed to beat a duration-matched public health factsheet.[6-7] Longer conversational agent strategies have begun to influence behavior change. For example, in a two-week school-based intervention, a chatbot delivered to parents tripled HPV vaccine uptake or appointment scheduling among their adolescent children compared to no intervention.[8] Together, these mixed findings leave open two core questions for clinical implementation: How much interaction with AI is actually necessary, and do shorter, one-off AI-generated messages already achieve most of the persuasive benefit?

Motivational interviewing (MI) offers a mechanistic lens on these questions. MI is a collaborative, person-centered counseling style that boosts motivation by helping individuals articulate their own reasons for change. In colorectal cancer screening, MI can address ambivalence, reinforce health values, and support informed choices. A 2022 systematic review of 14 randomized trials confirmed MI's superiority over usual care yet highlighted variability in effect sizes and delivery formats.[9] Additionally, MI can be labor intensive and hard to scale. LLMs could potentially emulate MI techniques and make them widely accessible.[10] Yet, it remains unclear if MI's empathetic, autonomy-supportive nature requires multi-turn dialogue or if a single, tailored AI message can convey enough relational depth to nudge intentions.

To examine these questions, we conducted a four-arm randomized controlled trial among 915 unscreened U.S. adults aged 45–75. Participants were randomized to receive no message, expert-written patient education material, a single demographically tailored GPT-4.1 message, or an MI-style chatbot conversation using GPT-4.1. All interventions were standardized for length and minimum exposure time to isolate the persuasive contribution of conversational depth. The primary outcomes were intent to complete stool testing or colonoscopy within 12 months, measured on 0–100 scales.

## METHODS

This four-arm randomized trial was approved by the University of Pennsylvania IRB (Protocol # 24-1977). The research was conducted per the Declaration of Helsinki and all participants provided informed consent. The protocol and analysis plan were preregistered (https://aspredicted.org/j8c5-388z.pdf) on 2025-05-19 and the study was retrospectively registered at ClinicalTrials.gov (NCT07107152) on 2025-07-30. Data were collected between May and June 2025.

U.S. adults aged 45–75 who reported never (or being unsure about) previously completing colorectal cancer (CRC) screening were recruited via an online research panel and randomized to: (1) no-message control (neutral reading), (2) expert-written patient material, (3) a single AI message, or (4) a motivational-interviewing (MI)-style AI chatbot. All participants were required to engage with the assigned stimulus for at least 3 minutes. The expert material was taken from The Journal of the American Medical Association's patient-oriented summary of the USPSTF 2021 screening recommendations; the AI conditions used GPT-4.1, with the chatbot prompted to employ MI techniques. Both AI conditions tailored content using participant's self-reported demographics including age and gender. Implementation details, full prompts, and example outputs appear in the Appendix (Tables S1-S2).

Because intervention formats were readily distinguishable, participants could not be blinded to treatment format. Participants assigned to the expert-written and single-AI-message conditions were not informed of the source of the material before exposure, whereas participants assigned to the chatbot condition were informed that they would interact with an AI system. Randomization and intervention delivery occurred automatically through the online study platform without investigator interaction. Outcomes were entered directly by participants in the online survey.

Primary outcomes were post-intervention intentions to complete (a) a colonoscopy and (b) a stool test within 12 months, each on 0–100 scales; baseline intentions were measured pre-intervention and included as covariates. Secondary outcomes included perceived empathy and message effectiveness and comfort discussing CRC screening, plus ratings of screening-hesitancy reasons. Participants failing attention checks or flagged by bot detection were excluded (criteria in Appendix).

Analyses used linear regression with HC2 robust standard errors, adjusting for baseline values. Standardized effect sizes (Cohen's d) were computed from model residual SDs; 95% CIs used the delta method. Prespecified sensitivity analyses with demographic/clinical covariates and exploratory moderation appear in the Appendix. As an exploratory sensitivity analysis for attrition, we repeated the primary baseline-adjusted linear regressions using baseline observation carried forward for participants who had baseline measures but did not complete the post-intervention assessment. For these participants, post-intervention colonoscopy and stool-test intent were set equal to baseline intent, corresponding to an assumption of no change.

As an exploratory robustness analysis, we also modeled the pre-to-post change score for each outcome as a function of study arm with HC2 robust standard errors. We also conducted post-hoc exploratory two one-sided tests (TOST) comparing the chatbot with the single-message condition separately for colonoscopy intent and stool-test intent. The equivalence margin was set at ±0.20 times the outcome-specific residual standard deviation from the primary ANCOVA model, corresponding to standardized bounds of -0.20 and 0.20. We considered equivalence to be established only if both one-sided p values were below .05 and the entire 90% confidence interval fell within the stated equivalence margin. As a post-hoc exploratory analysis among participants with baseline intent ≤50, we fit a logistic regression (HC2 robust SEs) estimating the odds of crossing >50 post-intervention; sensitivity analyses used 25 and 75 thresholds (details in Appendix).

We also conducted brief mechanistic and interpretive analyses including manual coding of open-text rationales, MI-behavior coding of chatbot turns, content analysis of single-message persuasion strategies, readability assessment, and clinician review of AI outputs (full procedures in Appendix).

The prespecified power analysis targeted 1,096 participants. Recruitment was stopped before reaching this target because of increasing difficulty identifying additional U.S. adults aged 45–75 on the online panel who met the eligibility criterion of never having completed CRC screening.

## RESULTS

Of 1,299 eligible participants invited to the study, 984 entered the baseline cohort, below the prespecified target of 1,096 because of difficulty recruiting additional adults who had never completed CRC screening. After 45 participants did not complete the post-intervention assessment and 24 failed an attention check, the primary analytic sample included 915 participants who reported never undergoing screening (colonoscopy, stool test, etc.) (see Figure 1 for CONSORT diagram). The sample was primarily female (63.7%), had a mean age of 51.6 years, and majority White (83.4%) (Table 1). The mean baseline intent to complete screening within 12 months was 30.9 (SD = 34.4) for a colonoscopy and 32.2 (SD = 32.8) for a stool test (out of 100).

Nearly all 695 participants assigned to the non-chatbot arms self-reported reading the assigned material. Specifically, 95.3% (221/232) of the expert material participants, 96.3% (208/216) of the single AI message, and 99.2% (245/247) of the no-message participants (assigned a fictional short story) reported reading carefully. One of the single AI message participants self-reported not reading the material, while the remaining 20 across arms reported skimming. Participants in the chatbot condition sent an average of 88.2 (SD = 69.0) total words, and the chatbot sent an average of 1226.2 (SD = 781.6) words (Appendix Table S3). The average chatbot message had a Flesch-Kincaid grade level of 9.1 (1.1), the average participant response had a grade level of 3.7 (2.0), the average single AI message had a grade level of 9.4 (1.1) and the expert-written material had a grade level of 10.8.

To assess factual accuracy and clinical appropriateness, two licensed clinicians each reviewed a separate random sample of 50 LLM generated messages: 25 chatbot responses and 25 single AI messages. Overall, all 100 responses were determined to be factually accurate and clinically appropriate.

### Primary Outcomes

Across all participants, intent to receive a stool test increased by an average of 8.9 (SD=22.9) points and increased by an average of 3.3 (SD=17.7) for colonoscopy intent (Appendix Figure S1). Overall, all message formats outperformed the no-message control in boosting stool test and colonoscopy intent. However, relative to the expert material, the AI conditions only boosted stool test intent, showing no significant effect for improving colonoscopy intent (Figure 2, Appendix Table S4).

To aid interpretation, we computed robust, baseline-adjusted Cohen's d for each pairwise contrast using the model residual SD (colonoscopy model $\sigma = 17.0$, stool test model $\sigma = 21.5$). Both the chatbot and single AI message formats produced medium effects on stool-test intent (ds = 0.64 and 0.60, respectively; both $ps < .001$), outperforming the expert materials ($d = 0.35$). Differences between the AI formats and expert material were statistically significant, though small. The chatbot-versus-single-message contrast was not statistically significant. Full pairwise comparisons and colonoscopy-related effects are presented in Appendix Table S5.

These effects remained robust to demographic controls, with estimated treatment effects similar in magnitude and direction to the primary specification (Appendix Table S6). Exploratory tests of moderation identified a small set of significant interactions including trust in the CDC, NIH, as well as prior discussions about CRC screening with participants' primary care providers (Appendix Table S7-S8). For example, for both colonoscopy and stool test intent, the chatbot intervention was more effective among participants with lower trust in the CDC, and among those who had not previously discussed CRC screening with their doctor.

In the exploratory TOST analysis, the colonoscopy margin was plus or minus 3.41 points (0.20 residual SD), and the 90% equivalence CI was -4.52 to 1.89 (one-sided p values, .141 and .008). The stool-test margin was plus or minus 4.30 points, and the 90% equivalence CI was -3.05 to 4.82 (one-sided p values, .015 and .077) (Table S9). Therefore, because the confidence interval extended beyond the lower equivalence and upper equivalence boundaries, respectively, equivalence was not established between the two AI-formats for either colonoscopy or stool-test intent. Exploratory change-score models produced results broadly consistent in direction with the primary ANCOVA. The AI-versus-expert estimates were 5.06 and 4.21 points for stool-test intent and -1.73 and -0.14 points for colonoscopy intent (Table S10).

Among those participants whose baseline intent ≤ 50, both AI formats markedly shifted intent beyond the 50-point threshold relative to the no-message control (Appendix Table S11). The chatbot increased the odds 4.4-fold for colonoscopy and 5.4-fold for stool testing, and the single-message AI raised them by roughly 4.4- and 5.3-fold, respectively (all $p < .01$). The expert patient material produced a more modest effect (OR ≈ 2.3 for colonoscopy, 2.5 for stool; stool $p < .05$, colonoscopy n.s.). These findings remained consistent when using alternate binary thresholds of 25 and 75 on the intention scale, with treatment effects of similar magnitude and direction across models (Appendix Table S11).

**Attrition and Sensitivity Analyses**

Forty-five participants who completed baseline measures but did not complete the post-intervention assessment were included in a baseline-observation-carried-forward sensitivity analysis. This yielded a sensitivity sample of 960 participants: 233 in the chatbot arm, 233 in the single-AI-message arm, 242 in the expert-material arm, and 252 in the control arm. Missing post-intervention outcomes occurred for 13/233 (5.6%), 17/233 (7.3%), 10/242 (4.1%), and 5/252 (2.0%) participants, respectively. The 24 completers who failed an attention check remained excluded from this analysis. Attrition differed modestly across arms ($\chi^2[3]=8.25$, $p=.041$). The sensitivity analysis yielded effect estimates similar in magnitude and direction to the primary analysis, suggesting that differential attrition was unlikely to account for observed treatment effects (Appendix Table S12).

**Secondary Outcomes**

Across 13 potential reasons for hesitancy scored on a 0-100 importance scale, we observed no changes across any intervention relative to the control condition, after applying Benjamini-Hochberg correction for multiple comparisons (Appendix Figure S2).

Comparing AI conditions to the expert material, participants found both more empathetic (Appendix Table S13). Only the single AI message was rated more highly than the expert material on a message effectiveness scale. Both AI conditions increased self-reported comfort discussing CRC screening with one's doctor relative to the no-message control and expert material (Appendix Table S13). Notably, participants spent an average of 180 seconds more with the chatbot than with the single AI message, for an average of 393 seconds (120% more time than the 3 minute requirement) (Table S14).

The three most common reasons for why the single AI message was convincing were cited as motivation and reassurance (34 participants), personalization and relevance (19 participants), and accuracy and informational quality (17 participants). The three most common response reasons for why the chatbot was convincing were cited as human-like interaction and quality (26 participants), accuracy and informational quality (22 participants), and communication style and language (14 participants). The three most common response reasons for why the expert patient material was not convincing or inadequate were cited as a lack of need (21 participants), missing or inadequate information (17 participants), and content and style critiques (13 participants).

**Chatbot Conversations**

Conversation coding with BOLT, a computational framework for quantifying MI behaviors in LLM-mediated dialogues, showed the chatbot relied most heavily on concrete elaboration and didactic content.[11] Across all dialogues, an average of 80% of turns evoked concrete elaboration, 65% provided teaching, and 51% included normalizing statements (Table 2). Less frequent but still present were problem-solving moves (30%), explicit identification of user needs (25%), and emotion check-ins (22%). Planning language and explicit discussion of emotions or conflict were relatively uncommon (<16 % of turns), and perspective-taking prompts appeared in only about 1% of turns. This pattern suggests the chatbot prioritized clarifying participants' concrete barriers and supplying factual information over deeper exploration of values or perspectives.

**Single AI Message Persuasion Strategies**

Content analysis of the single AI messages identified 49 distinct persuasion tactics, which we manually clustered into six conceptual themes: (i) facts & information, (ii) practical/process guidance, (iii) emotional appeals, (iv) identity-based appeals, (v) social-norm cues, and (vi) rhetorical style (Table 3, Appendix Table S15). GPT-4.1 labeling on a four-point prevalence scale (0 = none, 3 = high) showed every message employed a direct-address rhetorical style (mean = 3.00, sd = 0.0), making it the only universally applied tactic. After rhetorical themes, social-norm cues were most prominent (mean = 2.06), driven chiefly by references to family responsibility and peer role-modeling. Identity appeals ranked next (mean = 1.49), with empowerment, age appropriateness, and personal responsibility appearing most prominently, followed by factual information(mean = 1.41) and emotional appeals (M = 1.35). Practical/process guidance was least frequent overall (M = 1.27), though providing clear next steps was also prominent. In addition, cost or insurance information, analogies, and testimonials from survivors appeared rarely (<0.5 on the four-point scale). These patterns suggest the model prioritized motivational language, normative pressure, and high-level evidence cues over granular logistical or financial details when constructing these single persuasive pitches.

## DISCUSSION

This randomized trial found both a demographically tailored AI generated message and an interactive MI chatbot significantly increased intent to complete colorectal cancer screening relative to no intervention. However, when compared to expert-written educational materials, the AI interventions showed no added benefit for colonoscopy intent and led to significantly greater gains only in stool test intent. Although stool-based screening is less commonly used than colonoscopy in the United States, it remains a guideline-recommended CRC screening strategy.[12] These findings suggest a concise AI-generated message may be an effective and scalable approach to promoting preventive health behaviors, particularly for less utilized and less invasive screening options.

Prior research on LLMs and persuasive messaging has largely focused on either static, one-shot messages or multi-turn conversations in isolation.[5-7,13-17] Our study fills a critical gap by directly comparing these two formats in a controlled setting, holding personalization and minimum exposure time constant. We did not detect a statistically significant difference between the longer, interactive chatbot and the simpler single AI message, despite participants choosing to spend substantially more time engaging with the chatbot. However, exploratory equivalence testing using a ±0.20 SD margin did not establish equivalence between the two AI formats for either colonoscopy or stool-test intent. Therefore, our findings cannot be interpreted as demonstrating that the interventions were equally effective or that conversational depth provides no additional value. Rather, in this study we did not detect an advantage of our particular MI-based conversational intervention over a concise, demographically tailored message.

MI is well-supported as an effective approach to promote CRC screening, and prior work has found LLMs can emulate MI techniques well.[18] Yet, studies have also found LLM-based MI therapists more closely resemble low-quality human therapists than high quality therapists in their usage of MI techniques.[11] Our analysis of the MI chatbot's messages using a previously established framework indicated the majority of chatbot responses emphasized factual information and concrete elaboration over other techniques such as empathetic engagement. These limitations may help explain why we did not detect an advantage of the chatbot over the single message and highlight opportunities for improving MI emulation in AI systems. Nonetheless, given the substantially lower time and implementation burden of single AI messages and the absence of a detected chatbot advantage in this trial, static personalized messages may represent an operationally attractive approach for health systems seeking to promote screening uptake.

These effects, while modest in absolute terms, may have meaningful population-level implications. Among participants with low baseline intent (≤50 on a 0–100 scale), those who received a single AI message had 4.4 times greater odds of crossing the 50-point threshold (i.e., indicating that screening was more likely than not) than those in the no-message control. As an illustrative, back-of-the-envelope calculation, if the probability of crossing this threshold in the absence of the intervention were assumed to be 20%, applying the observed odds ratio would correspond to approximately 3,200 additional individuals crossing the threshold per 10,000 low-intent adults exposed to the intervention. This calculation serves to only illustrate the potential population scale of changes in stated intention. However, we cannot say how such changes translate into completed screening.

These findings may have practical implications for health-system outreach. Short, demographically tailored messages generated by LLMs can be integrated into patient portals, text-based reminders, or printed with mailed screening kits. The relative ease of deployment and low-time burden for patients suggest these messages could be used at scale in health systems seeking to close screening gaps, particularly for newer or less invasive modalities. Moreover, although the messages in this study were based on GPT-4.1 and not fine-tuned or manually edited, all outputs reviewed were deemed factually and clinically appropriate by a licensed clinician, further supporting their potential for clinical integration. The AI messages were written at approximately a ninth-grade reading level, above the American Medical Association's recommendation of sixth grade, and could likely be simplified in future iterations to better meet health literacy needs and improve patient engagement.[19] Rather than relying on real-time generation, health systems could pre-generate a library of messages tailored to key demographic profiles, validate them for accuracy and appropriateness, and deploy them repeatedly at scale across outreach channels.

Whether higher-quality conversational agents can provide clinically meaningful advantages over static personalized messages remains unresolved. Improvements in emotional responsiveness, values engagement, reflective listening, and adaptive planning may increase their persuasive effectiveness. Future studies adequately powered for direct comparisons between active interventions should test whether conversational systems that more faithfully implement MI, provide logistical support, or integrate behavioral tools such as appointment scheduling can outperform well-designed single messages.

Interestingly, the advantage of AI over expert materials was limited to stool testing and did not extend to colonoscopy. This pattern mirrors findings from a recent HPV vaccine chatbot trial, where LLM chatbots failed to outperform expert-written materials for boosting vaccine intent.[7] We hypothesize AI may be most persuasive for less familiar or lower-barrier decisions, where information gaps are greater and resistance is lower. Stool tests are often less well known to patients, creating opportunities for persuasive messaging to fill informational gaps.[20] In contrast, colonoscopies, while a more established and recommended screening option, may carry entrenched beliefs about discomfort or logistical challenges that are less amenable to change through brief interventions. These differences were observed despite similar baseline levels of intent across the two modalities, suggesting the disparity is unlikely driven by starting preferences or ceiling effects alone.

This study has several limitations. First, the sample was drawn from an online convenience panel and skewed toward White and female participants, which may limit generalizability to more diverse populations. Participation required internet access and comfort with digital surveys, which may select for higher digital literacy than would be observed in routine outreach. The findings therefore require validation in more diverse communities and safety-net populations and in pragmatic workflows that identify unscreened patients from clinical records and measure completed screening. Second, the study relied on self-reported intentions and our findings cannot establish an effect on completed screening behavior. However, intent has been shown to correlate with actual CRC screening.[21,22] For example, one prospective study in England found that among 922 participants who intended to undergo a flexible sigmoidoscopy and whose attendance could be verified, 79.9% subsequently attended screening.[23] Third, participants were required to engage with the assigned intervention for a minimum of three minutes, potentially overestimating the impact of static materials compared to naturalistic exposure. Fourth, several of the targeting variables used in the study may be infeasible to access in real-world implementation. However, additional variables such as race and zip code are typically accessible via electronic health records. Fifth, because the intervention format was apparent and outcomes were self-reported immediately after exposure, expectancy and demand characteristics may have contributed to the observed effects. This concern is especially relevant to comparisons with the neutral-reading control. However, comparisons with the expert-written material provide a more credible estimate of incremental benefit from AI-generated content. Sixth, the study was not powered or originally designed to establish equivalence between the chatbot and single-message interventions, and recruitment stopped below the prespecified target sample size due to recruitment difficulties. Accordingly, the absence of a statistically significant difference between the two AI formats cannot be interpreted as evidence of equivalent effectiveness. Lastly, although the primary outcomes and analysis plan were pre-registered, due to an oversight, Clinical Trial registration was completed retrospectively.

In conclusion, demographically tailored messages generated by LLMs represent a promising, low-touch approach to increasing CRC screening intent, particularly for stool-based tests. Given their substantially lower time and implementation burden, brief personalized messages may offer an operationally attractive strategy for health-system outreach over conversational chatbots. As health systems seek scalable and personalized strategies to address preventive care gaps, incorporating LLM-generated messages into outreach efforts may provide a practical, impactful tool for promoting evidence-based screening behaviors.

**Data availability**

The full, anonymized dataset and code used in this is available at https://github.com/sehgal-neil/CRC_LLM

**Acknowledgments**

This project was funded in part by the University of Pennsylvania's Penn Medicine Communication Research Institute, Penn Global Research and Engagement Fund, and National Institutes of Health, NIH-NIMHD:R01MD018340 awarded to Dr. Guntuku. The funders had no role in the design and conduct of the study; collection, management, analysis, and interpretation of the data; preparation, review, or approval of the manuscript; and decision to submit the manuscript for publication.

**Author Contributions**
Mr. Sehgal had full access to all of the data in the study and takes responsibility for the integrity of the data and the accuracy of the data analysis.
Concept and design: All authors
Acquisition, analysis, or interpretation of data: All authors
Drafting of the manuscript: NS
Critical revision of the manuscript for important intellectual content: All authors
Statistical analysis: NS, SCG
Obtained funding: SCG, AKA
Supervision: SCG

**Competing Interest Statement**
SJM serves on the Clinical Advisory Board of Guardant Health.

## Figures and Tables

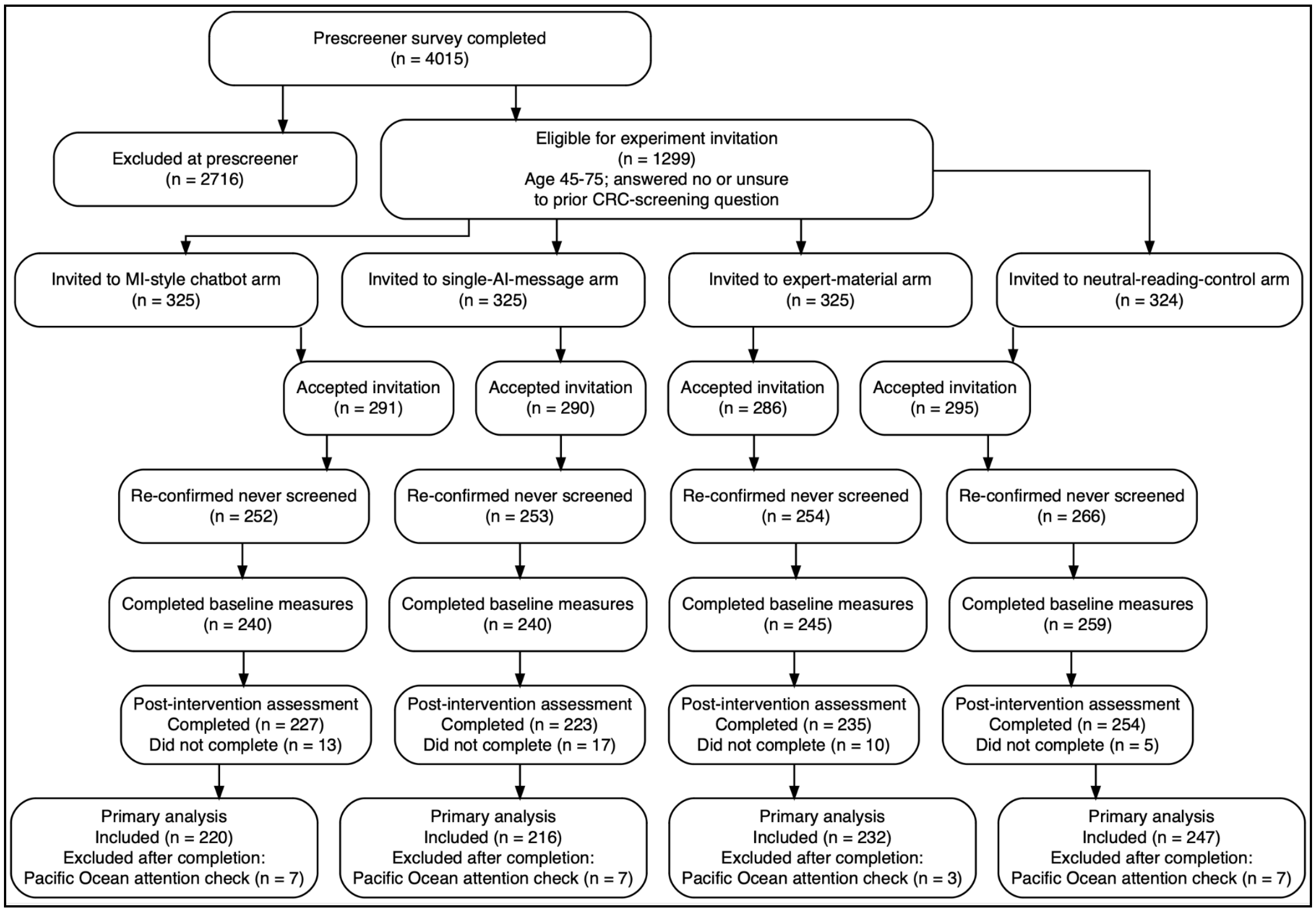

**Fig. 1. CONSORT Diagram of Participants**

A)

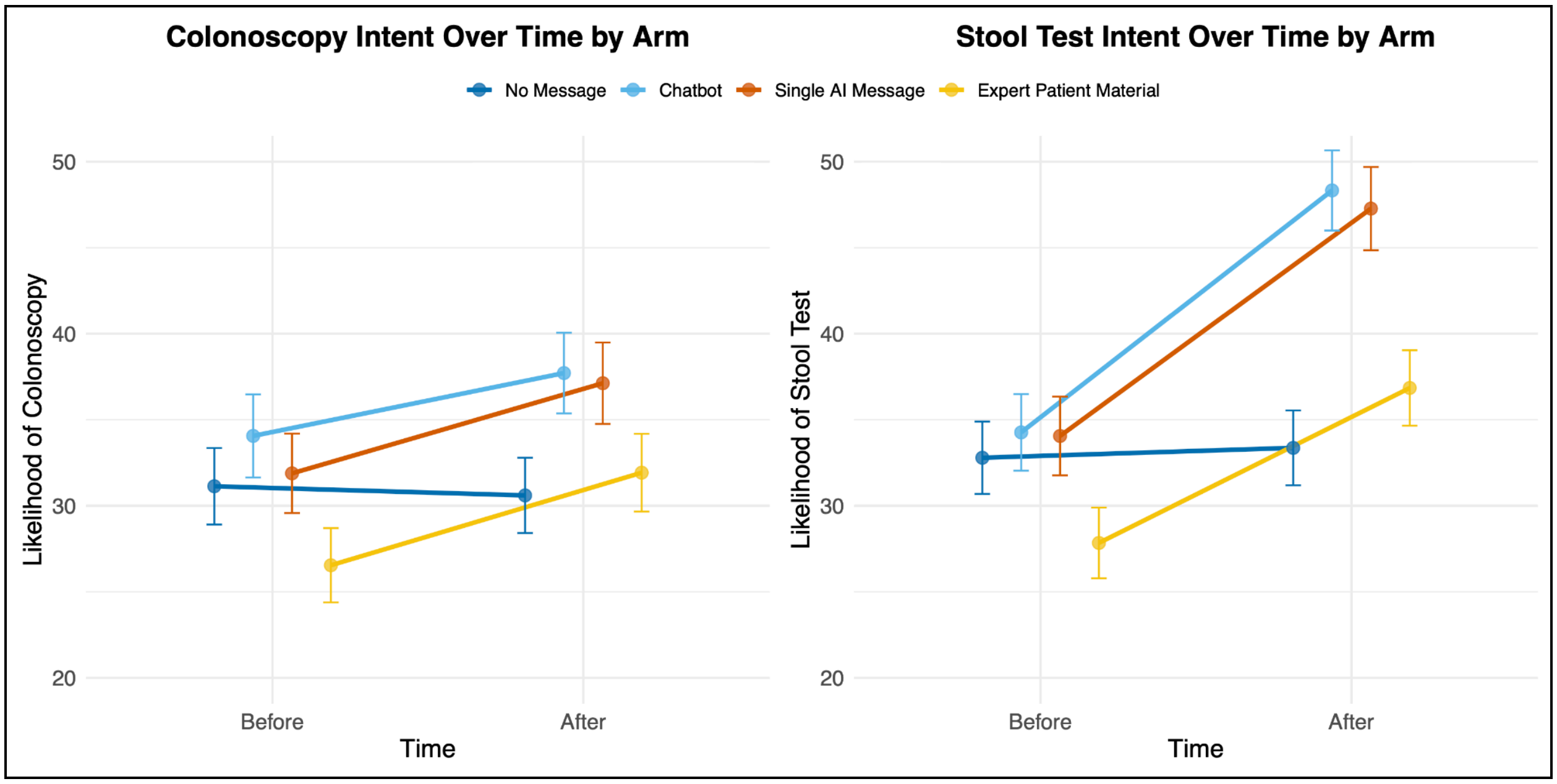
Colonoscopy Intent Over Time by Arm
Stool Test Intent Over Time by Arm
No Message
Chatbot
Single AI Message
Expert Patient Material
Likelihood of Colonoscopy
Likelihood of Stool Test
50
40
30
20
Before
After
Time

B)

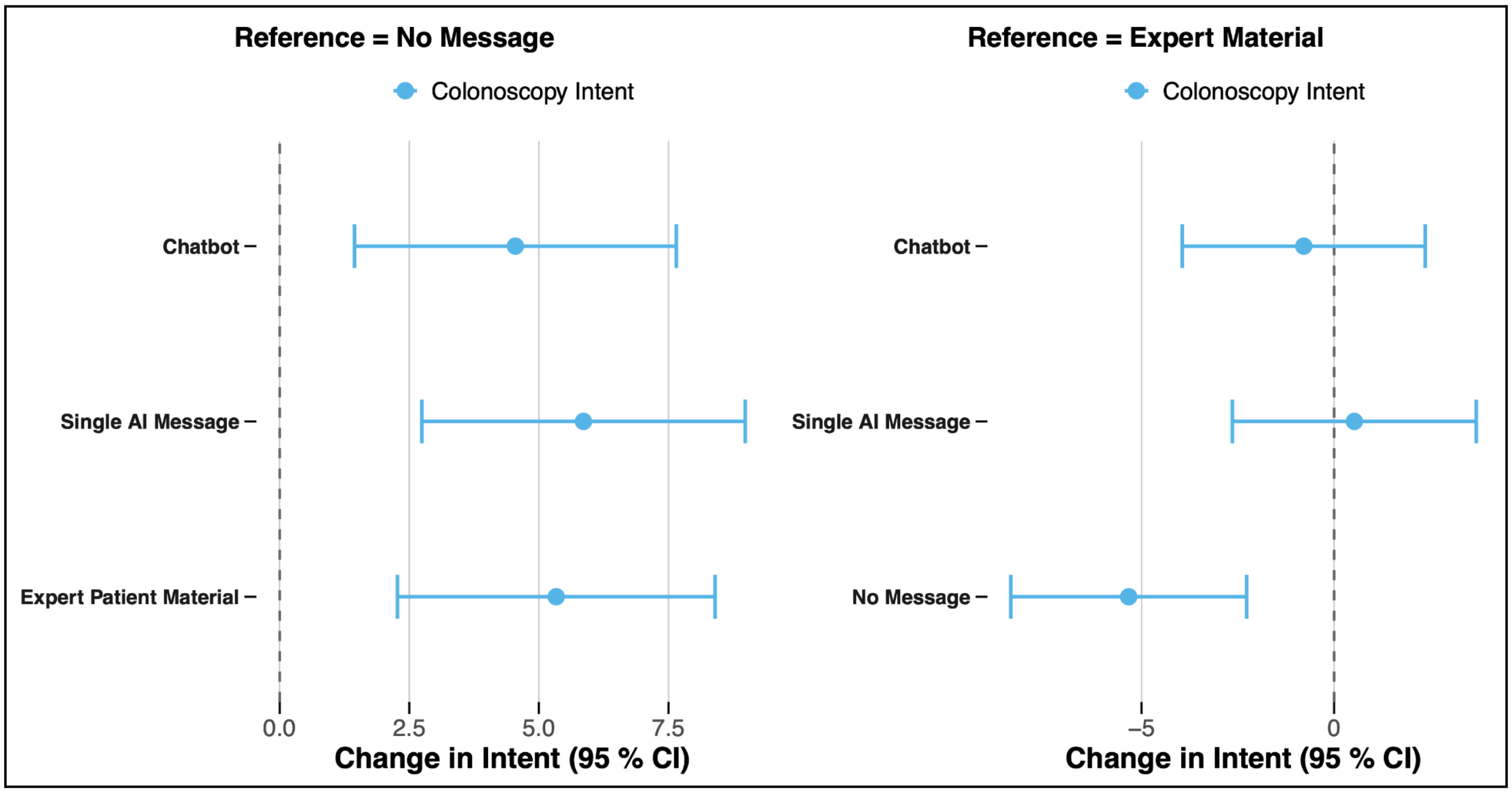
Reference = No Message
Reference = Expert Material
Colonoscopy Intent
Chatbot
Single AI Message
Expert Patient Material
No Message
0.0
2.5
5.0
7.5
−5
0
Change in Intent (95 % CI)

C)

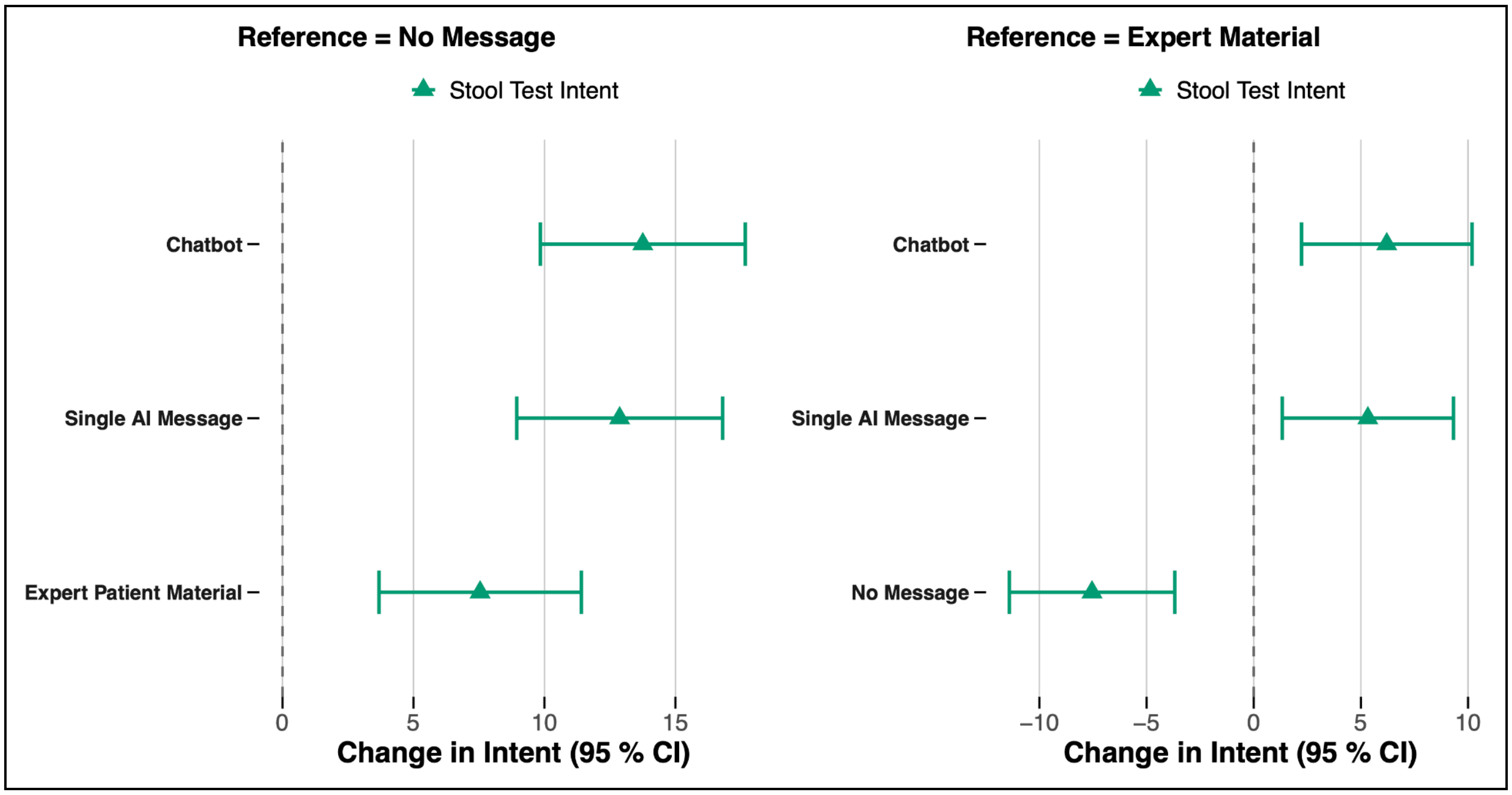

**Figure 2. Change in participant's intent to complete CRC screening within 12 months (N = 915)**

Panel A displays raw mean intention (± SE) for each arm at baseline and post-treatment. Vertical jitter is added for visibility and does not reflect a meaningful variable. Panel B displays HC2-robust 95% confidence intervals for increase in colonoscopy intent relative to the no message and expert patient material, derived from the OLS model in Table S4. Panel C displays HC2-robust 95% confidence intervals for increase in stool test intent relative to the no message and expert patient material, derived from the OLS model in Table S4. All three interventions significantly increased both colonoscopy and stool test intention relative to the No-Message control.

**Table 1.** Participant Characteristics by Arm
The randomized arms' observed baseline intent values are reported descriptively rather than used as a balance test.

| | Overall | No-Message | Chatbot | Single AI Message | Expert Material | p |
|---|---|---|---|---|---|---|
| **n** | 915 | 247 | 220 | 216 | 232 | |
| **Age (mean (SD))** | 51.61 (6.50) | 51.54 (5.94) | 51.11 (6.51) | 51.65 (6.45) | 52.13 (7.09) | 0.425 |
| **Gender** | | | | | | 0.842 |
| Man | 330 (36.1) | 93 (37.7) | 75 (34.1) | 76 (35.2) | 86 (37.1) | |
| Other | 2 (0.2) | 1 (0.4) | 0 (0.0) | 0 (0.0) | 1 (0.4) | |
| Woman | 583 (63.7) | 153 (61.9) | 145 (65.9) | 140 (64.8) | 145 (62.5) | |
| **Race** | | | | | | 0.665 |
| White | 763 (83.4) | 199 (80.6) | 189 (85.9) | 183 (84.7) | 192 (82.8) | |
| Black | 83 (9.1) | 25 (10.1) | 16 (7.3) | 21 (9.7) | 21 (9.1) | |
| Other | 69 (7.5) | 23 (9.3) | 15 (6.8) | 12 (5.6) | 19 (8.2) | |
| **Education** | | | | | | 0.008 |
| High School or Less | 109 (11.9) | 39 (15.8) | 19 (8.6) | 17 (7.9) | 34 (14.7) | |
| Some College | 357 (39.0) | 81 (32.8) | 77 (35.0) | 100 (46.3) | 99 (42.7) | |
| Bachelors Degree | 310 (33.9) | 89 (36.0) | 83 (37.7) | 71 (32.9) | 67 (28.9) | |
| Graduate Degree | 139 (15.2) | 38 (15.4) | 41 (18.6) | 28 (13.0) | 32 (13.8) | |
| **Politics** | | | | | | 0.443 |
| Very Conservative | 54 (5.9) | 20 (8.1) | 8 (3.6) | 17 (7.9) | 9 (3.9) | |
| Conservative | 130 (14.2) | 32 (13.0) | 32 (14.5) | 32 (14.8) | 34 (14.7) | |
| Somewhat Conservative | 110 (12.0) | 26 (10.5) | 28 (12.7) | 25 (11.6) | 31 (13.4) | |
| Moderate | 215 (23.5) | 62 (25.1) | 48 (21.8) | 44 (20.4) | 61 (26.3) | |
| Liberal | 184 (20.1) | 52 (21.1) | 47 (21.4) | 48 (22.2) | 37 (15.9) | |
| Somewhat Liberal | 99 (10.8) | 20 (8.1) | 25 (11.4) | 28 (13.0) | 26 (11.2) | |
| Very Liberal | 123 (13.4) | 35 (14.2) | 32 (14.5) | 22 (10.2) | 34 (14.7) | |
| **Urbanicity** | | | | | | 0.921 |
| Rural | 215 (23.5) | 56 (22.7) | 48 (21.8) | 50 (23.1) | 61 (26.3) | |
| Suburb | 482 (52.7) | 128 (51.8) | 119 (54.1) | 117 (54.2) | 118 (50.9) | |
| Urban | 218 (23.8) | 63 (25.5) | 53 (24.1) | 49 (22.7) | 53 (22.8) | |
| **Self Reported Health Status** | | | | | | 0.499 |
| Poor | 30 (3.3) | 7 (2.8) | 9 (4.1) | 6 (2.8) | 8 (3.4) | |
| Fair | 244 (26.7) | 59 (23.9) | 56 (25.5) | 58 (26.9) | 71 (30.6) | |
| Good | 537 (58.7) | 149 (60.3) | 130 (59.1) | 122 (56.5) | 136 (58.6) | |
| Excellent | 104 (11.4) | 32 (13.0) | 25 (11.4) | 30 (13.9) | 17 (7.3) | |
| **Trust in AI** | | | | | | 0.347 |
| Not at all | 63 (6.9) | 20 (8.1) | 10 (4.5) | 14 (6.5) | 19 (8.2) | |
| Slightly | 190 (20.8) | 56 (22.7) | 49 (22.3) | 45 (20.8) | 40 (17.2) | |
| Somewhat | 493 (53.9) | 137 (55.5) | 118 (53.6) | 107 (49.5) | 131 (56.5) | |
| Very much | 161 (17.6) | 33 (13.4) | 42 (19.1) | 47 (21.8) | 39 (16.8) | |
| Completely | 8 (0.9) | 1 (0.4) | 1 (0.5) | 3 (1.4) | 3 (1.3) | |
| **Trust in CDC (mean (SD))** | 3.47 (1.21) | 3.40 (1.26) | 3.55 (1.19) | 3.46 (1.25) | 3.47 (1.14) | 0.641 |
| **Trust in NIH (mean (SD))** | 3.49 (1.20) | 3.43 (1.24) | 3.57 (1.19) | 3.50 (1.22) | 3.44 (1.14) | 0.581 |
| **Trust in PCP (mean (SD))** | 3.84 (0.91) | 3.74 (0.98) | 3.89 (0.88) | 3.92 (0.86) | 3.84 (0.91) | 0.154 |
| **Last Seen PCP** | | | | | | 0.471 |
| In the past 6 months | 314 (34.3) | 75 (30.4) | 81 (36.8) | 88 (40.7) | 70 (30.2) | |
| In the past 12 months | 171 (18.7) | 46 (18.6) | 46 (20.9) | 35 (16.2) | 44 (19.0) | |
| In the past 2 years | 113 (12.3) | 32 (13.0) | 28 (12.7) | 22 (10.2) | 31 (13.4) | |
| In the past 5 years | 101 (11.0) | 28 (11.3) | 22 (10.0) | 21 (9.7) | 30 (12.9) | |
| More than 5 years ago | 216 (23.6) | 66 (26.7) | 43 (19.5) | 50 (23.1) | 57 (24.6) | |
| **Ever talked to doctor about CRC screening** | | | | | | 0.109 |
| No | 622 (68.0) | 168 (68.0) | 139 (63.2) | 151 (69.9) | 164 (70.7) | |
| Unsure | 15 (1.6) | 6 (2.4) | 0 (0.0) | 5 (2.3) | 4 (1.7) | |
| Yes | 278 (30.4) | 73 (29.6) | 81 (36.8) | 60 (27.8) | 64 (27.6) | |
| **Baseline Stool Test Intent (0-100)** | 32.2 (32.8) | 32.8 (33.0) | 34.3 (33.0) | 34.1 (33.6) | 27.8 (31.3) | – |
| **Baseline Colonoscopy Intent (0-100)** | 30.9 (34.4) | 31.1 (34.9) | 34.1 (35.8) | 31.9 (33.9) | 26.5 (32.9) | – |

**Table 2.** Chatbot Motivational Interviewing Technique Usage
Percentage of turns containing the technique, averaged across all chatbot dialogues (N = 220 conversations).

| Technique | Mean % of Turns |
|---|---|
| Evokes Concrete Elaboration | 80.04% |
| Teaching | 64.99% |
| Normalizing | 51.38% |
| Problem-Solving | 29.46% |
| Makes Needs Explicit | 24.89% |
| Makes Strengths/Resources Explicit | 22.82% |
| Emotions Check-in | 22.49% |
| Planning | 15.57% |
| Makes Emotions Explicit | 12.49% |
| Makes Conflict Explicit | 6.41% |
| Makes Values Explicit | 4.43% |
| Makes Consequences Explicit | 3.12% |
| Evokes Perspective Elaboration | 1.29% |

**Table 3.** Single AI Message Persuasion Theme Usage

| Theme | Mean (SD) Usage (0 - low, 3 - high) |
|---|---|
| Rhetorical | 3.00 (0.00) |
| Social Norms | 2.06 (0.38) |
| Identity Appeals | 1.49 (0.18) |
| Facts and Information | 1.41 (0.18) |
| Emotional Appeals | 1.35 (0.23) |
| Practical/Process | 1.27 (0.18) |

# APPENDIX

## EXTENDED METHODOLOGY

All study protocols and procedures were approved by the University of Pennsylvania's Institutional Review Board; all participants provided informed consent. All study hypotheses and analysis plans were preregistered at https://aspredicted.org/j8c5-388z.pdf and retrospectively registered at Clinicaltrials.gov (NCT07107152). All data were collected in Summer 2025.

### Participants
Adult participants residing in the United States were recruited through CloudResearch Connect, an online platform known to provide high-quality respondent samples for academic research.[1] Eligibility was determined via a prescreening survey, which identified adults aged 45-75 who reported never (or being unsure about) having undergone colorectal cancer (CRC) screening via colonoscopy, stool test, sigmoidoscopy, or CT colonography. Survey participation was compensated at a rate equivalent to 18 USD/hour. Participants who passed the eligibility criteria were randomized to one of four arms and invited to the main experiment survey consisting of baseline measures, the assigned intervention, and post-intervention measures. Randomization was performed in Python by generating a random allocation of Prolific IDs, which was then uploaded to the Prolific platform. Simple randomization without blocking was used.

### Baseline Measures
The experiment survey began with a baseline assessment that reconfirmed eligibility and collected a range of participant characteristics. These included demographic information such as age, gender, education level, political ideology, residential urbanicity, and self-reported health status. Participants also reported their engagement with the healthcare system, including the frequency of visits to a primary care provider (PCP) and whether they had previously discussed CRC screening with a physician. Trust in AI and health-related entities including the CDC, NIH, personal PCP was additionally collected. Intentions to undergo CRC screening were measured via self-reported likelihood of completing a colonoscopy and a stool test within the next 12 months, each on a 0 to 100 scale where "Extremely Unlikely" and 100 was marked "Extremely Likely". Participants also rated their comfort discussing CRC screening with a provider on a 0 to 100 scale. Reasons for CRC screening hesitancy were assessed using a series of items adapted from prior research on screening barriers; each was rated on a 0 to 100 scale, with 0 labeled "Not important at all" and 100 labeled "Extremely important".[2-3] An open-ended "Other" option allowed participants to report additional concerns. One attention check item was included to screen out inattentive responses.

### Intervention
All participants were presented with a stimulus and required to spend at least three minutes reading or engaging with the stimulus. A three-minute session length was chosen to mirror realistic, practical constraints for public-health use cases.

In the expert patient material condition, participants read material from The Journal of the American Medical Association's Patient Page on colorectal cancer screening which was a 642 word educational summary of the US Preventive Services Task Force's 2021 recommendations on screening for colorectal cancer.[4] Participants were not informed of the source prior to reading.

In the single AI message condition, participants read a single message written by GPT-4.1. GPT-4.1 was instructed to write a persuasive 642 word message for a pamphlet designed to persuade someone to get screened for colorectal cancer. The model was provided with each participant's age, gender, education level, political leanings, residential urbanicity, self-reported health status, and time since last PCP visit, as well as the full text of the JAMA Patient Page. Participants were not informed prior to reading the material that the source was an AI model.

In the no message control condition, participants read a neutral, 642 word fictional short story about a cat and dog during a storm.

In the chatbot condition, participants were informed they would engage in a brief conversation with a neutral, non-judgmental AI about a complicated topic. This interaction was implemented as a multi-turn exchange with GPT-4.1. The model was prompted to use motivational interviewing (MI) techniques to encourage CRC screening. The model was provided the same participant-specific characteristics as in the single message condition, as well as a 10-step MI roadmap adapted from prior research.[5]

Both the chatbot and single-message conditions were implemented using OpenAI's Chat-Completions API, embedded in a Qualtrics Vegapunk widget. No model fine-tuning, scripted logic, or additional training data were used. The two AI arms differed only in the format and system prompts used. We use OpenAI's best performing non-reasoning LLM at the time of the experiment, GPT-4.1, in order to maximize the potential persuasiveness of the AI conditions.[6] All model parameters were left at the default. Full prompt texts and example outputs are available in Appendix Tables S1-S2.

**Post-Intervention**

Immediately following the intervention, participants completed a series of outcome measures. The two primary outcome were the self-reported likelihood of (1) undergoing a colonoscopy and (2) completing a stool test within the next 12 months. Each was assessed separately on a 0-100 scale where 0 was marked "Extremely unlikely" and 100 was marked "Extremely likely". Secondary outcomes included the perceived empathy of the material or interaction on a 5 point scale, perceived effectiveness of the material/conversation (adapted from an existing public health message effectiveness scale[7], and comfort discussing CRC screening with a provider (0-100 scale). Participants were also asked to re-rate the reasons for CRC screening hesitancy, using the same items and format as in the baseline assessment. An additional attention check item was included to detect post-intervention inattentiveness. Lastly, participants were asked to write one to two sentences describing why they found their given intervention convincing and one to two sentences why their given intervention was not convincing. A priori sample size calculations using G*Power indicated that 1,096 participants were required to detect a small effect ($f=0.10$) with 80% power and $\alpha=0.05$.

**Analyses**

Participants who failed either of the two attention checks or received a bot detection score below 0.5 based on an embedded invisible reCAPTCHA were excluded from the analytic sample.[8]

Linear regressions with HC2 robust standard errors were used to estimate treatment effects. For outcomes with baseline measures, pre-intervention scores were included as covariates. Standardized effect sizes (Cohen's d) were obtained by dividing each HC2-robust coefficient by the model's pooled residual SD, with 95% confidence intervals obtained via the delta method using the same robust covariance matrix.

For robustness, we additionally estimated the primary treatment effect on CRC screening intentions using a linear regression that included the full set of measured individual characteristics: age (centered), gender (female vs. non-female), race (White vs. non-White), political ideology (liberal, conservative, moderate), urbanicity, self-reported health (good or excellent health vs. not), last time seeing PCP (within the past 2 years vs. not), prior CRC screening discussions with a doctor, trust in AI (very much or completely vs not), and trust in the NIH, CDC, and personal PCP. To explore potential moderators, we estimated separate regression models interacting the treatment variable with each individual characteristic, applying Benjamini-Hochberg correction for multiple comparisons.

To further explore intervention effects among participants with initially low intent, we conducted a post-hoc exploratory analysis limited to individuals with baseline screening intent ≤ 50. We fit a logistic regression model predicting whether participants crossed the 50-point threshold post-intervention, using study arm as the predictor and HC2-robust standard errors to compute odds ratios with 95% confidence intervals. As a robustness check, we additionally fit models examining 25 point and 75 point thresholds.

To understand the reasons behind participants' persuasive responses, one co-author manually analyzed open-text explanations provided by participants who found the intervention convincing. This analysis was limited to responses from individuals whose post-treatment intention scores increased for at least one primary outcome and exceeded a value of 50. Conversely, open-text responses from participants who did not find the intervention convincing were also manually annotated. These included only those whose post-treatment intention scores were equal to or lower than their corresponding pre-treatment scores. All thematic categories derived from these manual annotations were subsequently reviewed and validated by a second co-author.

To characterize the chatbot exchanges, we applied BOLT, a computational framework for quantifying motivational-interviewing (MI) behaviors in LLM-mediated dialogues.[9] Each chatbot-generated turn was parsed through BOLT's in-context classifiers, which tag 13 canonical MI moves (e.g., open questions, affirmations, reflections). We then normalized the counts by the number of turns to determine the normalized frequency of each code within a conversation and derived descriptive distributions for the full chatbot corpus.

Persuasion strategies in the single AI message arm were examined with a hybrid deductive–inductive content analysis used in prior work on AI-generated messaging.[10] First, GPT-4.1 was prompted ten times to list strategies that could persuade an unscreened adult to complete colorectal cancer screening. After manual de-duplication, one author iteratively coded a random set of messages, adding new categories until no additional tactics emerged, producing a 49-item list. Finally, GPT-4.1 was supplied with the finalized strategy list and asked to rate each strategy's presence in every message on a four-point ordinal scale (none, low, moderate, high). We then categorized each strategy into one of five manually curated themes for interpretation.

To understand the level of complexity across message formats, we computed the grade level required to read each of the chatbot conversations, single AI messages, and expert material using the Flesch-Kincaid algorithm.

To assess factual accuracy and clinical appropriateness, two licensed clinicians each reviewed a separate set of 25 random chatbot conversations and 25 random single AI messages.

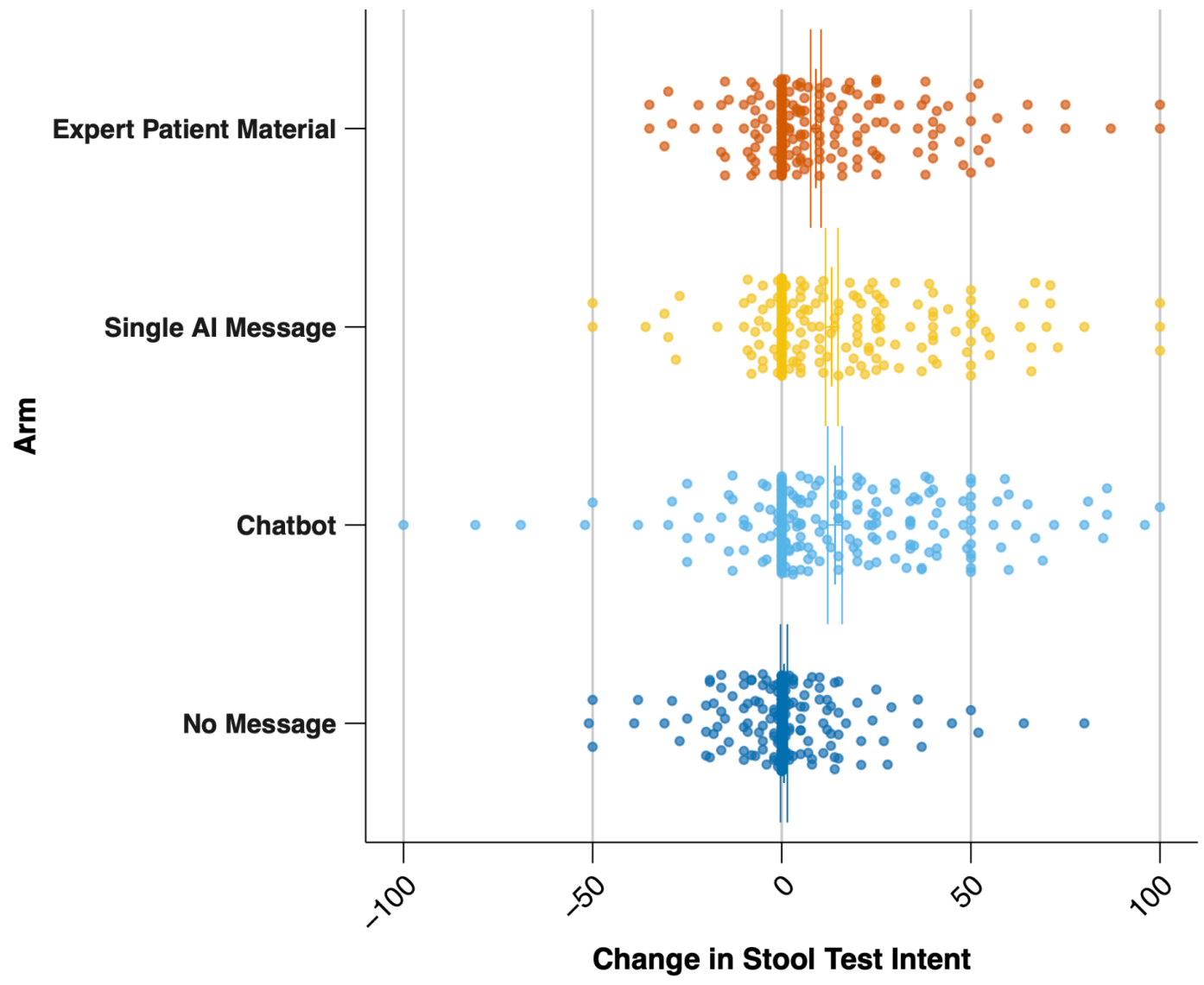

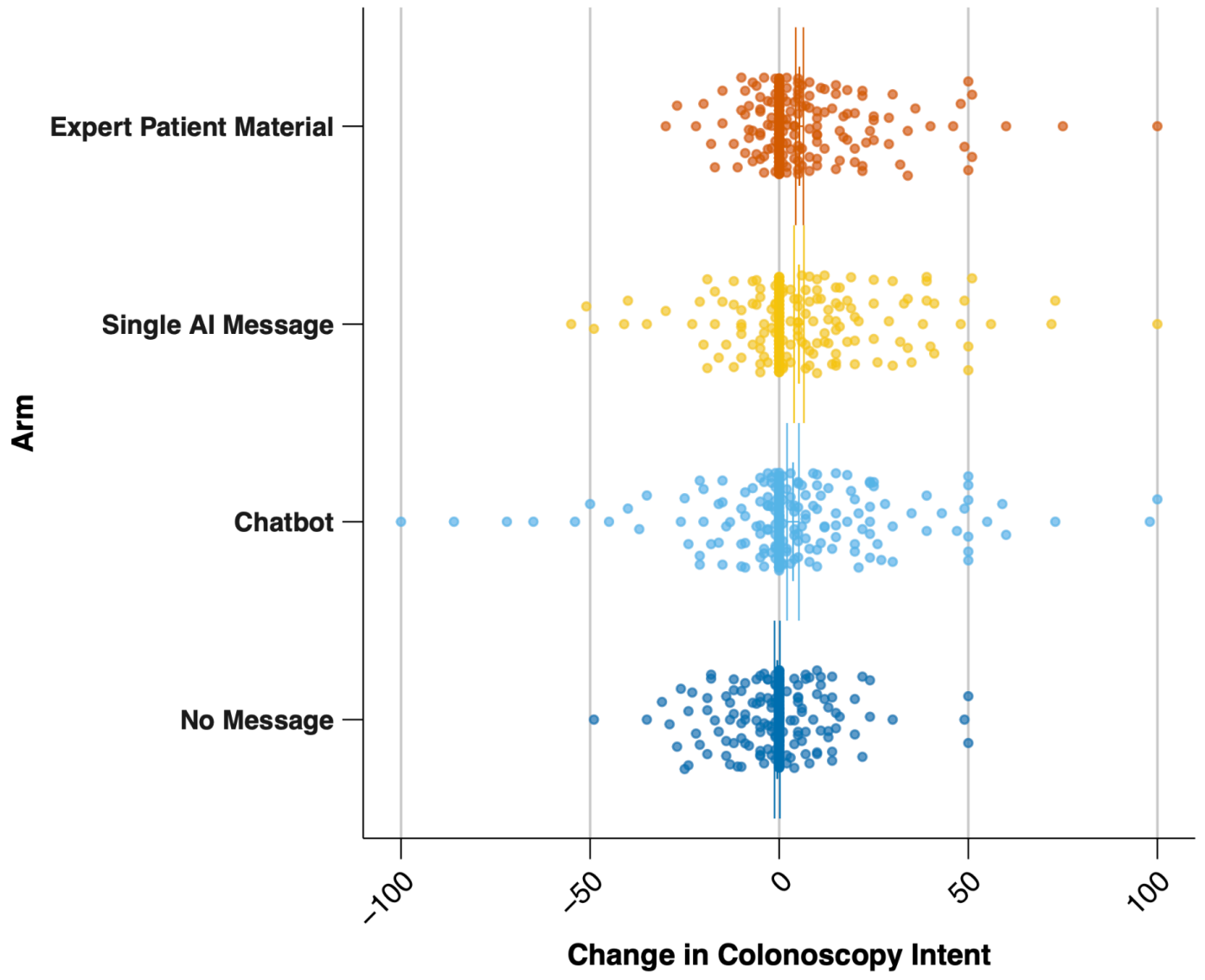

**Fig. S1.** Participant Level change in intention for each arm at baseline and post-treatment. Vertical lines display mean and standard errors. Horizontal jitter is added for visibility and does not reflect a meaningful variable.

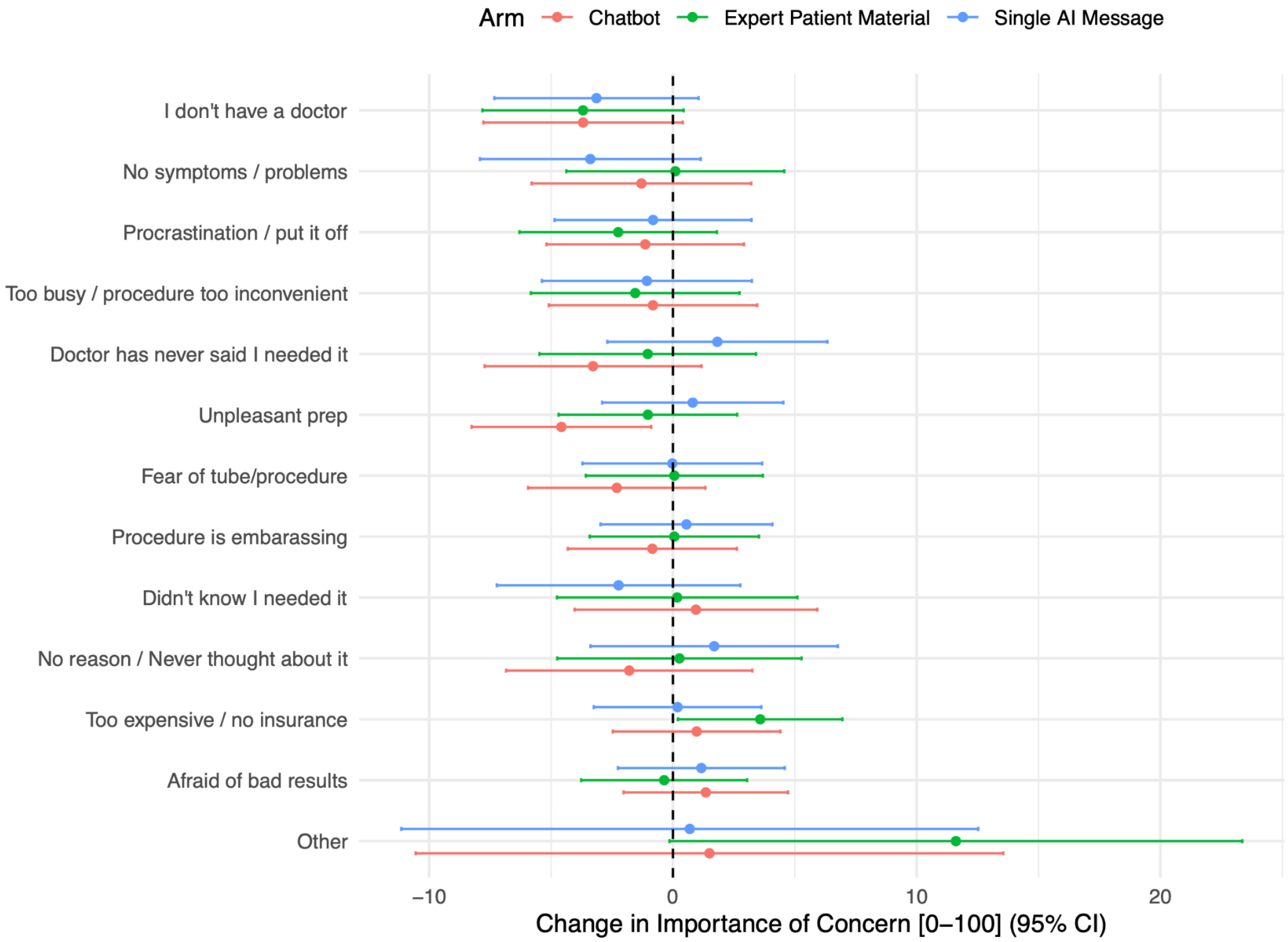

**Fig. S2.** Intervention effects on 13 concerns for CRC screening hesitancy
Coefficient plot (point estimates ± 95 % CI) from separate OLS models for importance of specific concerns (scale 0: Extremely Unimportant to 100: Extremely Important); estimates are relative to the No-Message control. After applying Benjamini-Hochberg correction for multiple comparisons, no changes are significant relative to the no message control. Concerns were not captured for overall screening hesitancy, and were not elicited separately for colonoscopies and stool tests.

**Table S1.** LLM Prompts

| Prompt | Model | Prompt |
|---|---|---|
| **Chatbot Prompt** | GPT-4.1 | "You will be talking to a person who has not been screened for colorectal cancer. Your goal is to effectively persuade them to get screened for colorectal cancer by using motivational interviewing techniques. Below is a 10-step roadmap for how to do motivational interviewing.\n\n## Table of Contents\n1. [Purpose & Preparation]\n2. [Introduction & Engagement]\n3. [Agenda Setting]\n4. [Eliciting the Patient's Perspective]\n5. [Reflective Listening & Summaries]\n6. [Providing Information (Elicit–Provide–Elicit)]\n7. [Exploring Ambivalence & Building Motivation]\n8. [Addressing Barriers & Building Confidence]\n9. [Inviting the Patient to Plan Next Steps]\n10. [Closing & Follow-Up]\n11. [Final Notes and Tips]\n\n---\n\n## 1. Purpose & Preparation\n\n**Goal:** Encourage colorectal cancer screening through a collaborative, empathetic conversation that respects the patient's autonomy and experiences.\n\n1. **Review Patient Chart**  \n   - Note age, medical/family history, prior screening history, and any known barriers (financial, logistical, psychosocial).\n\n2. **Plan the Setting**  \n   - Ensure a quiet, private environment.  \n   - Sit at eye level; use a welcoming posture.\n\n---\n\n## 2. Introduction & Engagement\n\n**Goal:** Establish rapport, trust, and a comfortable atmosphere for open communication.\n\n1. **Greeting & Warm Welcome**  \n   - "Hello [Patient Name], it's good to see you today. How are you feeling?"\n\n2. **Explain Your Role & Intent**  \n   - "I'd like to talk about your health and ways we can help you stay well. One area is screening for colorectal cancer, which is recommended for people [within your age range/with your history]."\n\n3. **Express Empathy & Interest**  \n   - "I understand that talking about cancer screening can feel overwhelming or unpleasant. I want to hear your thoughts and experiences."\n\n---\n\n## 3. Agenda Setting\n\n**Goal:** Collaboratively define what will be discussed and invite the patient's input.\n\n1. **Outline Topics**  \n   - "I'd like to talk about your health goals and how colorectal cancer screening might fit in. But first, I'd like to hear your perspective and any questions or concerns you may have."\n\n2. **Ask for Patient Input**  \n   - "Is there anything specific you want to make sure we discuss about screening today?"\n\n3. **Confirm & Collaborate**  \n   - "So, we'll focus on learning about where you stand on screening, what questions you have, and what steps—if any—you'd like to take next. Does that sound good?"\n\n---\n\n## 4. Eliciting the Patient's Perspective\n\n**Goal:** Use open-ended questions to learn about the patient's thoughts, feelings, concerns, and knowledge around CRC screening.\n\n1. **Open-Ended Questions**  \n   - "What do you already know about colorectal cancer screening?"  \n   - "What comes to mind when you think about having a colonoscopy or a stool test?"\n\n2. **Explore Feelings/Emotions**  \n   - "How do you feel about the idea of screening for colon cancer?"  \n   - "What worries or concerns do you have about it?"\n\n3. **Listen Without Interruption**  \n   - Give the patient space to express themselves freely.\n\n4. **Identify Ambivalence**  \n   - If the patient expresses mixed feelings, acknowledge it:  \n     - "It sounds like part of you |
| | | wants the protection that screening offers, but you also feel uneasy about the procedure.”\n\n---\n\n## 5. Reflective Listening & Summaries\n\n**Goal:** Demonstrate empathy and clarify understanding of the patient’s concerns. Reflective listening helps patients hear their own motivations and ambivalence in a non-judgmental space.\n\n1. **Reflective Listening (Simple Reflection)**  \n   - **Patient:** “I’m worried it might be painful.”  \n   - **Provider (reflection):** “You’re worried that the procedure might be uncomfortable or hurt.”\n\n2. **Reflective Listening (Complex Reflection)**  \n   - **Patient:** “I just keep putting it off because it doesn’t feel urgent and I’m busy.”  \n   - **Provider (reflection):** “It sounds like life has been hectic, and you’d rather focus on immediate responsibilities than a test that doesn’t feel pressing, even though you know it’s important.”\n\n3. **Summarizing**  \n   - “Let me see if I understand correctly: You know screening is recommended, and you’d like to prevent colon cancer. At the same time, you feel anxious about the test, you’re worried it might be painful, and you’re not sure if you can fit it into your schedule right now. Is that right?”\n\n---\n\n## 6. Providing Information (Elicit–Provide–Elicit)\n\n**Goal:** When offering information or advice, do so in a way that respects the patient’s autonomy and fosters collaboration.\n\n1. **Elicit**  \n   - “Would you like more information about the different screening tests and how they work?”  \n   - “Is it okay if I share some information about why screening is recommended?”\n\n2. **Provide**  \n   - Offer clear, concise information on:  \n     - **Reason for Screening:**  \n       - “Colorectal cancer often has no symptoms early on. Screening helps catch it early or even prevent it by finding and removing polyps.”  \n     - **Screening Options:**  \n       - “The most common are colonoscopy, which can be done every 10 years if normal, and stool-based tests, which can be done yearly or every three years, depending on the test.”  \n     - **Safety & Comfort Measures:**  \n       - “We use sedation for colonoscopy. Many people only experience mild discomfort.”\n\n3. **Elicit**  \n   - “What are your thoughts on this information?”  \n   - “How does this fit with your experiences or concerns?”\n\n---\n\n## 7. Exploring Ambivalence & Building Motivation\n\n**Goal:** Help the patient sort out their mixed feelings and gently guide them toward recognizing the importance of screening.\n\n1. **Decisional Balance**  \n   - “What do you see as the benefits of getting screened?”  \n   - “What might be some downsides or challenges you see with screening?”\n\n2. **Reflect & Summarize Ambivalence**  \n   - “It sounds like you want to reduce your risk of colon cancer, but the idea of the procedure makes you uneasy and you’re not sure how to fit it in.”\n\n3. **Use Scaling Questions**  \n   - “On a scale from 0 to 10, how important is it for you to do screening?”  \n   - “On a scale from 0 to 10, how confident are you that you could get screened if you decided to?”  \n   - **Follow-Up:**  \n     - “Why are you at a [current number] and not at a lower number?” (Helps elicit positive reasons or motivations.)\n\n---\n\n## 8. Addressing Barriers & Building Confidence\n\n**Goal:** Help the patient overcome barriers and boost their self-efficacy.\n\n1. **Brainstorm Solutions**  \n   - “Let’s talk about anything that could make screening feel easier or more doable for you.”\n\n2. **Support Autonomy & Strengths**  \n   - “You mentioned you value your health and staying active. Those |
| | | strengths can help you get through this test.”\n\n3. **Offer Practical Supports** \n - **Work/Scheduling:** “We can schedule the procedure at a time that’s convenient for you. Here are some options.” \n - **Cost:** “Let’s review your insurance coverage or explore any assistance programs.”\n\n---\n\n## 9. Inviting the Patient to Plan Next Steps\n\n**Goal:** Collaboratively form a clear plan that aligns with the patient’s values and readiness.\n\n1. **Recap & Invite** \n - “We’ve talked about why screening is important, your concerns, and possible ways to address them. What do you think would be your next step?”\n\n2. **Offer Choice & Self-Determination** \n - “Would you prefer to schedule a colonoscopy, or consider a stool test first? It’s entirely up to you.”\n\n3. **SMART Plan** \n - **Specific:** Which test (colonoscopy vs. fecal immunochemical test [FIT], etc.) \n - **Measurable:** Scheduling date or timeframe for an at-home test \n - **Achievable:** Confirm insurance coverage, sedation options, transportation help \n - **Relevant:** Ties back to preventing cancer, staying healthy \n - **Time-bound:** Set a date or timeline\n\n---\n\n## 10. Closing & Follow-Up\n\n**Goal:** Provide a supportive end to the conversation and ensure the patient knows the next steps.\n\n1. **Summarize Key Points** \n - “We decided that you’ll do the FIT test once a year, and we’ll schedule a reminder to see your results. If there’s anything abnormal, we’ll arrange a colonoscopy.”\n\n2. **Affirm the Patient’s Effort** \n - “I appreciate you sharing your thoughts with me today. Thank you for taking the time to consider screening.”\n\n3. **Plan for Follow-Up** \n - “I’ll check in with you in [X weeks/months] to see how it went and answer any questions. Does that work for you?”\n\n4. **Encourage Ongoing Communication** \n - “If any questions come up, or if something changes, please let me know. I’m here to help you.”\n\n---\n\n## Final Notes and Tips\n\n- **Stay Person-Centered:** Maintain empathy and support. The goal is not to pressure the patient, but to guide them toward an informed decision that feels right to them. \n- **Check for Understanding (Teach-Back):** Ensure patients understand the information by having them repeat it in their own words. \n- **Be Culturally Sensitive:** Adapt language, examples, and respect any relevant cultural or spiritual beliefs influencing the patient’s decision-making. \n- **Continue Using MI Skills:** Even when providing information, maintain open-ended questions, reflective listening, and summarizing to keep the patient engaged and empowered.\n\nThe person you will talk to has the following characteristics:\nage: " + responses.age + "\npolitical ideology: " + responses.politics + "\ngender: " + responses.gender + "\neducation: " + responses.education + "\ncommunity: " + responses.community + "\nself-reported health: " + responses.health + "\nlast time seeing PCP: " + responses.pcp + "\n\n\nThe conversation will now begin. You will be making the first move and introduce yourself as an AI assistant designed to help them learn about colorectal cancer screening options\n" |
| **Single AI Message Prompt** | GPT-4.1 | "Please write a persuasive 642 word message for a pamphlet designed to persuade someone to get screened for colorectal cancer\n\nThe person who will receive the pamphlet has the following following characteristics:\nage: " + responses.age + |
| | | "\npolitical ideology: " + responses.politics + "\ngender: " + responses.gender + "\neducation: " + responses.education + "\ncommunity: " + responses.urban_rural + "\nself-reported health: " + responses.health + "\nlast time seeing PCP: " + responses.pcp + "\n\nYou may not provide false information.\n\nBelow is some up to date information taken from JAMA: \n\nThe US Preventive Services Task Force (USPSTF) has recently published updated recommendations on screening for colorectal cancer.\n\nWhat Is Colorectal Cancer?\n\nColorectal cancer refers to cancer of either the colon (large intestine) or the rectum, which are the lower parts of the gastrointestinal tract. Colorectal cancer mostly affects older adults. However, in recent years, the rate of diagnosis in adults younger than 50 years has been increasing. Symptoms of colorectal cancer include changes in bowel habits, abdominal pain, and sometimes blood in the stool. Colorectal cancer can be treated by surgery, chemotherapy, or sometimes radiation therapy.\n\nWhat Tests Are Used to Screen for Colorectal Cancer?\n\nDirect visualization looks at the colon and rectum using a scope and camera (colonoscopy or flexible sigmoidoscopy) or by computed tomography (CT) scan (CT colonography, also called virtual colonoscopy). These tests require bowel preparation using laxatives to empty the colon and rectum beforehand. Stool-based tests involve collecting a sample of stool to look for either blood or abnormal genetic material, both of which can be signs of cancer. Examples include the high-sensitivity fecal occult blood test, fecal immunochemical test, and stool DNA test.\n\nThere are pros and cons to each of these tests, and there are different screening intervals recommended for each. The USPSTF does not specifically recommend using one test versus another.\n\nWhat Is the Patient Population Under Consideration for Screening for Colorectal Cancer?\n\nThis USPSTF recommendation applies to adults aged 45 years or older who do not have symptoms of colorectal cancer. It does not apply to people who are at higher risk of colorectal cancer, such as those who have a strong colorectal cancer family history, inflammatory bowel disease, or a history of colon polyps.\n\nWhat Are the Potential Benefits and Harms of Screening for Colorectal Cancer?\n\nScreening can prevent people from dying of colorectal cancer. Clinical trials show that screening adults aged 50 to 75 years decreases colorectal cancer deaths. Computer simulation studies suggest that starting screening at age 45 years may help some people live longer than if they started screening at age 50 years. For people aged 76 to 85 years, the benefit of screening is smaller. Among this older group, those who have never been screened are more likely to benefit. Harms of screening are primarily due to colonoscopy (either initial screening colonoscopy or colonoscopy done to follow up abnormal results of other screening tests). Uncommon but potential serious harms include bleeding or infection from the procedure.\n\nHow Strong Is the Recommendation to Screen for Colorectal Cancer?\n\nGiven the current evidence, the USPSTF concludes with high certainty that the net benefit of screening for colorectal cancer in adults aged 50 to 75 years is substantial. The USPSTF concludes with moderate certainty that the net benefit of screening for colorectal cancer in adults aged 45 to 49 years is moderate and that the net benefit of |

<table>
<tr><td></td><td></td><td>screening for colorectal cancer in adults aged 76 to 85 years who have been screened in the past is small. Adults who have never been screened for colorectal cancer are more likely to benefit."</td></tr>
<tr><td>Single AI Message Annotator</td><td>GPT-4.1</td><td>Evaluate the use of persuasive strategies in a colorectal-cancer-screening pamphlet.<br><br>Examine the pamphlet and rate how strongly each pre-defined persuasive strategy appears using the provided scale.<br><br>**RATING SCALE**:<br><br>- **"none"** – 0 explicit mentions, or only an indirect passing phrase<br>- **"low"** – 1 mention, brief (<20 words) and not reiterated or emphasized<br>- **"moderate"** – EITHER 2–3 distinct mentions, OR one extended sentence (20–40 words), OR highlighted once (e.g., heading/bullet)<br>- **"high"** – ≥4 mentions OR the strategy frames the pamphlet (e.g., appears in title/heading **and** conclusion, or the cumulative text devoted to it ≥80 words)<br><br>**Definition of "mention"**: A distinct sentence, bullet, heading, or clause that clearly conveys the strategy.<br><br># Output Format<br><br>Return only a single JSON object. Keys must match the exact name of each strategy listed below. Values should be one of the four rating strings: "none," "low," "moderate," or "high." Do not add markdown, explanations, or extra keys.<br><br>```json<br>{<br>  "Highlight Reduced Cancer Risk": "none | low | moderate | high",<br>  "Reference Doctor Recommendations": "none | low | moderate | high",<br>  "Highlight Endorsements from Trusted Organizations": "none | low | moderate | high",<br>  "Highlight Screening Simplicity": "none | low | moderate | high",<br>  "Reassure About Minimal Discomfort": "none | low | moderate | high",<br>  "Cite Screening Statistics": "none | low | moderate | high",<br>  "Share Statistics on Success": "none | low | moderate | high",<br>  "Normalize Routine Screening": "none | low | moderate | high",<br>  "Counter Screening Misconceptions": "none | low | moderate | high",<br>  "Address Common Fears": "none | low | moderate | high",<br>  "Emphasize Ease of Stool Tests": "none | low | moderate | high",<br>  "Emphasize Pros of Flexible Sigmoidoscopy": "none | low | moderate | high",<br>  "Provide Clear Next Steps": "none | low | moderate | high",<br>  "Share Testimonials from Survivors": "none | low | moderate | high",</td></tr>
</table>

<table>
<tr><td></td><td></td><td>"Emphasize Peace of Mind": "none | low | moderate | high",<br>"Reassure Privacy and Dignity": "none | low | moderate | high",<br>"Promote Free or Low-Cost Options": "none | low | moderate | high",<br>"Focus on Quick Recovery": "none | low | moderate | high",<br>"Reassure on Procedure Safety": "none | low | moderate | high",<br>"Counter Embarrassment Stigma": "none | low | moderate | high",<br>"Spotlight Local Screening Access": "none | low | moderate | high",<br>"Focus on Non-Invasive Choices": "none | low | moderate | high",<br>"Connect to Long-Term Health": "none | low | moderate | high",<br>"Mention Quick Results Timeline": "none | low | moderate | high",<br>"Present Easy Scheduling Options": "none | low | moderate | high",<br>"Appeal to Family Responsibility": "none | low | moderate | high",<br>"Explain Cost and Insurance Coverage Details": "none | low | moderate | high",<br>"Highlight One-Time or Infrequent Need": "none | low | moderate | high",<br>"Appeal to Community Norms or Peer Behavior": "none | low | moderate | high",<br>"Use of Direct Address": "none | low | moderate | high",<br>"Emphasize Empowerment and Personal Agency": "none | low | moderate | high",<br>"Highlight Role Modeling for Others": "none | low | moderate | high",<br>"Appeal to Social Justice or Health Equity": "none | low | moderate | high",<br>"Appeal to Science and Evidence-Based Action": "none | low | moderate | high",<br>"Appeal to Common Sense/Practicality": "none | low | moderate | high",<br>"Use of Analogies": "none | low | moderate | high",<br>"Appeal to Gender Roles": "none | low | moderate | high",<br>"Appeal to Self-Respect": "none | low | moderate | high",<br>"Appeal to Career Impact": "none | low | moderate | high",<br>"Appeal to Active Lifestyle": "none | low | moderate | high",<br>"Appeal to Education/Intelligence": "none | low | moderate | high",<br>"Appeal to Self-Identity": "none | low | moderate | high",<br>"Appeal to Age Appropriateness": "none | low | moderate | high",<br>"Appeal to Patriotism/National Identity": "none | low | moderate | high",<br>"Appeal to Milestones/Life Events": "none | low | moderate | high",<br>"Appeal to Independence/Self-Sufficiency/Personal Responsibility": "none | low | moderate | high",<br>"Appeal to Rural or Urban Identity/Challenges": "none | low | moderate | high",<br>"Appeal to Use of Earned Benefits": "none | low | moderate | high",<br>"Appeal to Enjoyment of Retirement/Leisure": "none | low | moderate | high",<br>}<br>```<br><br># Notes</td></tr>
</table>

| | | |
|---|---|---|
| | | - Carefully analyze each persuasive strategy for clarity and relevance.<br>- Use exact names and specified format for returning results without deviations. |

**Table S2.** LLM Output Examples
Participants saw text enclosed in one asterisk displayed in italics. Participants saw text enclosed in two asterisks displayed in bold.

| **Arm** | **Example Output** |
|---|---|
| Chatbot | Chatbot<br>Hello! My name is HealthAI, and I'm an assistant here to support you in staying healthy. One topic I'd like to help with is learning about ways to screen for colorectal cancer—a common, but largely preventable, disease. I'd like to make sure you have all the information you need about your screening options, and that you feel comfortable exploring any questions or concerns you might have. How does that sound?<br><br>Participant<br>Sounds good. Can I start with a question for you?<br><br>Chatbot<br>Absolutely! Please, feel free to ask any questions you have. I want this to be as helpful to you as possible. What's on your mind?<br><br>Participant<br>Great, what percentage of people get colorectal cancer?<br><br>Chatbot<br>That's a really important question, and I'm glad you asked.<br><br>Colorectal cancer is actually one of the most common cancers in the United States. About 1 in 23 men (roughly 4-5%) and 1 in 25 women (about 4%) will be diagnosed with colorectal cancer at some point in their lives.<br><br>Most cases occur after age 50, which is why screening becomes important around your age. The good news is that screening can help find precancerous polyps and catch cancer early, when it's most treatable.<br><br>How does that information sit with you? Would you like to know more about how to reduce your risk or about the screening process itself?<br><br>Participant<br>That's interesting. I have another question. Is it hereditary?<br><br>Chatbot<br>Great question. Colorectal cancer can have a hereditary component, but most cases are not directly inherited.<br><br>About 1 in 3 people who develop colorectal cancer have a family member who also had it—most commonly a parent, sibling, or child. There are specific genetic syndromes, like Lynch syndrome or familial adenomatous polyposis (FAP), that can significantly increase risk, but these are relatively rare.<br><br>For most people, the risk comes from a combination of age, lifestyle, and sometimes family history. Even for those without a family history, screening is very important because colorectal cancer can develop "out of the blue." |
| | Do you know if anyone in your family has had colorectal cancer or polyps? Or would you like to talk more about how your personal or family background factors into your own risk?<br><br>Participant<br>No one in my family has had it. What types of lifestyle choices increase the chances of getting it?<br><br>Chatbot<br>That's a thoughtful question, and it's great that you're thinking about factors you can control.<br><br>Several lifestyle choices can influence the risk of developing colorectal cancer:<br><br>- **Diet:** Eating a lot of red meat (like beef or pork) and processed meats (like hot dogs, bacon, or deli meats) can increase risk.<br>- **Physical activity:** Being inactive increases risk; regular physical activity helps lower it.<br>- **Weight:** Being overweight or obese is associated with higher risk.<br>- **Smoking:** Long-term smokers have a higher risk of colorectal cancer.<br>- **Alcohol:** Heavy, regular alcohol use can increase risk as well.<br><br>On the flip side, eating plenty of fruits, vegetables, and whole grains, staying active, maintaining a healthy weight, and avoiding tobacco are all linked to a lower risk.<br><br>Even with a healthy lifestyle, though, regular screening is important since early colorectal cancer often doesn't have symptoms.<br><br>Does anything about this surprise you, or is there a particular area you'd like to talk more about?<br><br>Participant<br>Honestly, I feel even more confident that I'm at very low risk of getting it. I don't drink or smoke. I'm not overweight, I exercise regularly and eat a healthy diet (with a bit too much candy mixed in!).<br><br>Chatbot<br>Thank you for sharing that, and it sounds like you've made a lot of positive choices for your health—those habits really do make a difference! It's completely understandable to feel reassured by your low-risk lifestyle.<br><br>At the same time, even people who do everything "right" can still develop colorectal cancer, especially because most cases happen after age 50, regardless of family history or lifestyle. That's one of the main reasons screening is recommended for everyone in your age group: it can find small changes or growths—like polyps—before they become a problem, even if you're feeling completely healthy.<br><br>I hear that you feel confident about your health, and you definitely have strong habits. How do you feel about the idea of screening when you don't have any symptoms and you're living a healthy lifestyle?<br><br>Participant<br>It wouldn't harm me, but it's also not something that I'm concerned about at all.<br><br>Chatbot |

That makes a lot of sense—you're healthy, feel well, and have no family history, so it's not at the top of your concerns. You also recognize that getting screened likely wouldn't hurt you, but it might not feel urgent or necessary.

Many people feel the same way, especially when they don't have any symptoms or risk factors. The challenge is that colorectal cancer often develops silently for years. Screening is one of the few ways to catch it when it's most treatable—or even prevent it entirely by removing precancerous polyps.

Would you be interested in hearing about the different screening options, including some that are simple and can be done at home? Or is there something specific that would make you more open to considering screening?

Participant
I'm open to any information. But, I'm also very likely to ignore it. While anything is possible, I'm pretty sure that it's not going to be cancer that kills me.

Chatbot
Thank you for being so open and honest. It's completely normal to feel that way—especially when you're feeling well and taking good care of yourself. It sounds like you value staying informed but don't feel a strong personal risk or urgency about colorectal cancer.

If it's okay, I'll share a little information just so you have it—you can use it now or anytime in the future. There are a few main screening options:

- **Colonoscopy:** Done just once every 10 years if results are normal. It both detects cancer and can prevent it by removing polyps.
- **Stool-based tests (like FIT or Cologuard):** Non-invasive, easy to do at home, and recommended every year or every three years, depending on the test. These look for tiny traces of blood or DNA changes.
- **Other options:** Like flexible sigmoidoscopy (less common).

Screening is recommended because, among healthy people, most cases of colorectal cancer can be caught early—or prevented altogether. Most people who are diagnosed with colorectal cancer thought they were "the last person it would happen to."

If you ever change your mind, know that screening is available and can be made as convenient as possible. What do you think about the home-based stool test option or scheduling a reminder to revisit this topic in the future, just in case?

And if you want, we can talk about any specific concerns, or how to keep screening stress-free. What would you like to do next—would you like more details, or just keep this information in your back pocket for now?

Participant
My doctor told me that I should get an exam a couple of years ago, but I just never got around to it.

Chatbot
Thank you for sharing that—it sounds like your doctor brought up screening before, and though you understood the recommendation, life just got busy and it wasn't a top priority. That's a really common experience; even when people intend to follow through, other responsibilities and routines can easily take over.

If you don’t mind my asking, what made it hard to get around to it before? Was it mostly about time, concerns about the test itself, or something else?

And if you decided you wanted to go ahead with screening, what do you think might help make it easier to follow through this time?

User
It's just the time combined with the fact that I'm very low risk. How long does an exam take?

Chatbot
That’s a great question—the time commitment can be a real barrier, especially with a busy schedule.

Here’s what to expect with the main options:

**Colonoscopy:**
- The actual procedure usually takes about 30 minutes.
- Most of your time is spent preparing: the day before, you’ll drink a special solution to clean out your colon and follow a specific diet for about 24 hours.
- After the procedure, you’ll need a little time to recover and someone to drive you home because of sedation.
- Most people return to normal activities by the next day.

**Stool-based tests (like the FIT test):**
- These are the easiest time-wise. You collect a small stool sample at home and send it to a lab.
- No prep, sedation, or recovery time needed.
- Just a few minutes to do the test whenever it fits your schedule.

For many people who are low risk and short on time, the stool test offers a way to get screened without disrupting their schedule.

How does hearing about the time involved with these options fit with your expectations? Would one of these approaches seem more doable for you?

Participant
Honestly, I'll probably not get a colonoscopy based upon what you wrote. I'd be more likely to take the FIT test. How would I get a kit?

Chatbot
That’s really helpful to hear, and I appreciate your honesty. Choosing the FIT test is a great, practical option for people who are busy and prefer the least disruption—plus, it still provides valuable screening.

To get a FIT kit, here’s what you typically do:

1. **Ask Your Doctor or Primary Care Provider:**
   - They can order a FIT kit for you, which is usually mailed to your home or picked up at the clinic.

2. **Receive the Kit:**
   - It comes with clear instructions. You collect a small stool sample—usually with a stick or brush—then put it in a provided container.

3. **Mail or Drop Off the Sample:**
   - You send it back to the lab in a prepaid envelope or drop it off at a designated location.

4. **Get Your Results:**
   - Your doctor’s office will let you know the results, usually within a week or two.

If the result is normal, you repeat the test every year. If anything unusual shows up, your doctor will discuss next steps (sometimes a colonoscopy).

Would you like help with how to request a FIT kit from your doctor, or would a reminder to bring it up at your next visit work better for you? What feels easiest for your next step?

Participant
I think I'll remember to bring it up the next time I have an appointment, although I'm also bad about seeing my PCP. It's just not something that I've ever prioritized because I've never had a health issue other than things like injuries.

Chatbot
That’s totally understandable. If you’ve always been healthy and only needed care for the occasional injury, regular check-ups or preventive screenings might not feel urgent or necessary. You’re certainly not alone in that—many people put off routine care when they feel well.

If you don’t mind me asking, would it feel helpful to schedule a preventive visit (even just once every year or two), or do you prefer continuing to make appointments only when needed? Some people find that putting it on the calendar, like any other maintenance task, helps keep their overall health on track—including opportunities to talk about things like the FIT test without needing to remember on the spot.

Regardless, it sounds like you have a pretty good plan: You’re open to bringing up the FIT test at your next appointment, and you’re aware that preventive care is available if you choose to prioritize it in the future.

Would a reminder from your clinic or a note to yourself help, or would you prefer to just leave it as something you’ll address when it comes up? I’m here to help make it as easy as possible, on your terms.

Participant
Last time I made an appointment, something came up and I cancelled it and never called back to reschedule. I just don't think it's that important, at least for me.

Chatbot
Thank you for your honesty—it sounds like you have a lot going on, and prioritizing preventive appointments hasn’t felt necessary for you, especially when you’re feeling healthy. Canceling and not rescheduling is something many people do, particularly when nothing feels urgent.

It’s clear you take care of yourself in ways that make sense for your life, and you value your time and priorities. At the same time, it sounds like you’re not completely ruling it out in the future; you’re just not putting it at the top of your to-do list right now.

If you ever do decide to make an appointment, it might help to know that preventive visits cover things beyond just screening—they’re also a chance to talk about any

| | |
|---|---|
| | health or wellness goals, get updated on any routine checks, and catch small issues before they become bigger problems.<br><br>I'm here to answer any questions or help you think through options whenever you need. Is there anything else about colorectal cancer screening, preventive visits, or your health that you'd like to talk about, or would you prefer to leave it here for now?<br><br>Participant<br>We can leave it there, and I appreciate the info that you shared with me!<br><br>Chatbot<br>Thank you for your openness today—I appreciate your willingness to talk about these things and I respect your approach to your health. If you ever have more questions about colorectal cancer screening, preventive care, or anything else, I'm always here as a resource.<br><br>Just remember, even small actions like asking about a FIT test at your next visit can make a big difference in staying healthy for years to come. And if life gets busy and it slips your mind, that's okay—you can always revisit it when the time feels right for you.<br><br>Take care, and thanks again for having this conversation! |
| Single AI Message | **Protect Your Health, Protect Your Family: Why It's Time to Get Screened for Colorectal Cancer at Age 48**<br><br>---<br><br>**A Vital Message for Men Who Take Responsibility Seriously**<br><br>As a 48-year-old man living in a suburban community, you know the value of responsibility. Whether it's providing for your family, caring for your home, or contributing to your neighborhood, you make choices every day that keep your life—and the lives of those you love—on track.<br><br>There's one more choice you can make right now that truly embodies what it means to be responsible: getting screened for colorectal cancer.<br><br>---<br><br>**Why Colorectal Cancer Should Be On Your Radar—Right Now**<br><br>You may feel healthy and not have any symptoms. But here's the reality: colorectal cancer doesn't always show signs until it's advanced. The rates of this cancer are rising among adults under 50, making screening more important than ever before.<br><br>Colorectal cancer refers to cancer of the colon or rectum. Common symptoms, like changes in your bowel habits or abdominal pain, often appear late—after the disease has had a chance to develop. This is why early detection is crucial: when detected early, colorectal cancer is one of the most treatable forms of cancer.<br><br>---<br><br>**The Trusted Recommendations—Backed by Science** |

The U.S. Preventive Services Task Force, a group of independent experts that reviews the strongest medical evidence, recommends that adults start screening for colorectal cancer at age 45. For healthy men like you, waiting any longer only increases the risk of missing the window when cancer can be treated most effectively—even prevented altogether.

Studies show that screening adults from ages 50 to 75 can reduce deaths from colorectal cancer significantly. Now, gathering even more evidence, experts agree: getting screened beginning at 45 could help save even more lives.

---

**What Are Your Options? Multiple Effective Tests, One Mission: Keeping You Healthy**

There’s no one-size-fits-all approach. You and your health provider can choose the screening method that’s right for you:

- **Colonoscopy:** A powerful tool that allows doctors to see the inside of your colon and remove abnormal growths before they can turn into cancer.
- **Stool-based tests:** Simple tests you can do at home, like the fecal immunochemical test or stool DNA test, catch early signs through a single stool sample.
- **CT Colonography (Virtual Colonoscopy):** An advanced scan that gives doctors a detailed image of your colon.

The goal isn’t to push you toward any specific test. The goal is to get you screened—your health and peace of mind are too important to leave unchecked.

---

**Addressing Concerns: Risks Are Real, But the Benefits Far Outweigh Them**

As a man who weighs every decision carefully, it’s smart to consider both the benefits and the risks. The main risks from screening—like bleeding or infection—are rare, and mostly linked to colonoscopy. For most men your age, the benefit of early detection and prevention overwhelms these uncommon risks. And remember, less invasive options are available.

---

**Conservative Values: Self-Reliance and Protecting What Matters Most**

Conservative principles are about more than politics—they’re about preserving what matters most: faith, family, freedom, and the ability to look out for yourself and your loved ones. Cancer, unfortunately, threatens all of that.

Getting screened is a proactive step—a decision that means you’re taking care of your health *before* there’s a problem. It keeps you in the driver’s seat, not cancer.

If you haven’t seen your primary care provider lately, now is the time to schedule a visit and talk about screening. You’ll be setting an example of strength and foresight for your family and your community.

---

| | |
|---|---|
| | **A Simple Choice That Could Save Your Life**<br><br>You work hard every day to protect the people and values you love. Protect yourself, too. Colorectal cancer screening is a practical choice that just makes sense.<br><br>**Talk to your doctor. Get screened. Stay strong for your family.**<br><br>*Take charge of your health today—because your community and your loved ones are counting on you.* |

**Table S3.** Chatbot Conversation Statistics

| | **Chatbot** | **Participants** |
|---|---|---|
| Turns | 8.4 (3.4) | 7.4 (3.3) |
| Words | 1226.2 (781.6) | 88.2 (69.0) |
| Words / Turn | 144.1 (41.5) | 12.3 (7.7) |
| Flesch-Kincaid Grade Level | 9.1 (1.1) | 3.7 (2.0) |

**Table S4.** Results from Ordinary Least Squares Regression (SEs Reported Below Coefficients)

| | Colonoscopy Intent (Reference = No message) | Colonoscopy Intent (Reference = Expert Material) | Stool Test Intent (Reference = No message) | Stool Test Intent (Reference = Expert Material) |
|---|---|---|---|---|
| | | | | |
| No Message | -- | -5.33*** | -- | -7.54*** |
| | | -1.22 | | -1.62 |
| | | | | |
| Chatbot | 4.55*** | -0.79 | 13.75*** | 6.21*** |
| | -1.64 | -1.78 | -2.02 | -2.23 |
| | | | | |
| Single AI Message | 5.86*** | 0.53 | 12.87*** | 5.32** |
| | -1.44 | -1.61 | -1.85 | -2.09 |
| | | | | |
| Expert Material | 5.33*** | -- | 7.54*** | -- |
| | -1.22 | | -1.62 | |
| | | | | |
| Pre-treatment Colonoscopy Intent | 0.87*** | 0.87*** | | |
| | -0.02 | -0.02 | | |
| | | | | |
| Pre-treatment Stool Test Intent | | | 0.82*** | 0.82*** |
| | | | -0.02 | -0.02 |
| | | | | |
| Constant | 3.37*** | 8.70*** | 6.44*** | 13.98*** |
| | -0.84 | -1.15 | -1.13 | -1.54 |
| | | | | |
| | | | | |
| Observations | 915 | 915 | 915 | 915 |
| R2 | 0.76 | 0.76 | 0.62 | 0.62 |
| Adjusted R2 | 0.76 | 0.76 | 0.62 | 0.62 |
| | | | | |
| Note: | | | | *p<0.1; **p<0.05; ***p<0.01 |

**Table S5.** Pairwise contrasts of message format effects on screening intent, reported as baseline-adjusted Cohen's d.
Positive values indicate higher intent for the first group in each contrast. Medium effects observed for chatbot and single AI message formats on stool-test intent; all colonoscopy effects were small. Full model used residual standard deviations of $\sigma = 17.0$ (colonoscopy) and $\sigma = 21.5$ (stool test).

| Outcome | Contrast | Cohen's d (SE) | p value |
|---|---|---|---|
| Colonoscopy | No Message - Chatbot | -0.27 (0.1) | <.01 |
| Colonoscopy | No Message - Single AI Message | -0.34 (0.09) | < 0.001 |
| Colonoscopy | No Message - Expert Material | -0.31 (0.07) | < 0.001 |
| Colonoscopy | Chatbot - Single AI Message | -0.08 (0.11) | 0.50 |
| Colonoscopy | Chatbot - Expert Material | -0.05 (0.11) | 0.66 |
| Colonoscopy | Single AI Message - Expert Material | 0.03 (0.09) | 0.74 |
| Stool test | No Message - Chatbot | -0.64 (0.09) | < 0.001 |
| Stool test | No Message - Single AI Message | -0.6 (0.09) | < 0.001 |
| Stool test | No Message - Expert Material | -0.35 (0.08) | < 0.001 |
| Stool test | Chatbot - Single AI Message | 0.04 (0.11) | 0.71 |
| Stool test | Chatbot - Expert Material | 0.29 (0.1) | <.01 |
| Stool test | Single AI Message - Expert Material | 0.25 (0.1) | 0.01 |

**Table S6.** Regression Results for Primary Outcome with Pre-Intervention Covariates (SEs Reported Below Coefficients)

| | Colonoscopy Intent (Reference = Expert Material) | Stool Test Intent (Reference = No message) |
|---|---|---|
| | | |
| Chatbot | 4.52*** | 13.50*** |
| | -1.64 | -2.03 |
| | | |
| Single AI Message | 5.63*** | 12.53*** |
| | -1.45 | -1.89 |
| | | |
| Expert Material | 5.35*** | 7.04*** |
| | -1.24 | -1.61 |
| | | |
| No Message | -- | -- |
| | | |
| | | |
| Age (centered) | -0.15* | 0.02 |
| | -0.09 | -0.12 |
| | | |
| Female Gender | 1.28 | 2.50* |
| | -1.18 | -1.47 |
| | | |
| White Race | -1.94 | -1.06 |
| | -1.63 | -1.91 |
| | | |
| Conservative (Ref: Liberal) | 2.37 | 1.83 |
| | -1.64 | -1.86 |
| | | |
| Moderate (Ref: Liberal) | 2.97** | 4.52** |
| | -1.38 | -1.85 |
| | | |
| Suburb (Ref: Rural) | 1.32 | -0.29 |
| | -1.39 | -1.7 |
| | | |
| Urban (Ref: Rural) | 0.64 | 0.24 |
| | -1.71 | -2.2 |
| | | |
| Good Self Reported Health | 0.21 | -2.69* |
| | -1.28 | -1.63 |
| | | |
| High Trust in AI | 2.65 | 2.8 |
| | -1.63 | -2.01 |
| | | |
| Trust in CDC | -1.32 | -1.52 |
| | -1.32 | -1.61 |
| | | |
| Trust in NIH | 1.09 | 0.75 |
| | -1.35 | -1.62 |
| | | |
| Trust in PCP | 0.48 | 1.68* |
| | -0.85 | -0.99 |
| | | |
| Seen PCP within 2 years | 0.43 | 2.56* |
| | -1.24 | -1.54 |
| | | |
| Never talked to doctor about CRC screening | 2.19 | 0.73 |
| | -1.39 | -1.78 |
| | | |
| Pre-treatment Colonoscopy Intent | 0.88*** | |
| | -0.02 | |
| | | |
| Pre-treatment Stool Test Intent | | 0.82*** |
| | | -0.02 |
| | | |
| Constant | -1.67 | 0.75 |
| | -4.16 | -4.7 |
| | | |
| | | |
| Observations | 915 | 915 |
| R2 | 0.76 | 0.63 |
| Adjusted R2 | 0.76 | 0.62 |
| | | |
| Note: | | *p<0.1; **p<0.05; ***p<0.01 |

**Table S7.** Colonoscopy Intent Moderator Analysis
OLS Regressions with Interaction. P-Values are BH corrected. SEs reported below coefficients.

| | Post-Treatment Colonoscopy Intent | | | | | | | | | | | |
|---|---|---|---|---|---|---|---|---|---|---|---|---|
| | | | | | | | | | | | | |
| Chatbot | 4.46** | 3.95 | 9.80 | -0.58 | 9.59** | 1.12 | 3.71 | 17.88*** | 12.47* | 18.36* | 1.99 | -2.27 |
| | -1.63 | -2.89 | -4.98 | -2.14 | -3.18 | -3.28 | -1.84 | -5.54 | -5.35 | -7.62 | -1.87 | -2.92 |
| | | | | | | | | | | | | |
| Single AI Message | 5.88*** | 4.55 | 5.11 | 4.77** | 5.96 | 3.54 | 5.30*** | 3.41 | 3.86 | -3.87 | 4.63** | 2.29 |
| | -1.44 | -2.49 | -3.36 | -1.93 | -3.01 | -2.82 | -1.51 | -3.55 | -4.02 | -6.16 | -1.74 | -2.60 |
| | | | | | | | | | | | | |
| Expert Material | 5.40*** | 5.45** | -0.08 | 3.34* | 4.23 | 5.79* | 5.09*** | -0.69 | -2.64 | -0.21 | 3.48* | 0.90 |
| | -1.23 | -1.77 | -2.80 | -1.59 | -2.06 | -2.48 | -1.26 | -3.64 | -3.44 | -6.25 | -1.46 | -2.08 |
| | | | | | | | | | | | | |
| No Message | -- | -- | -- | -- | -- | -- | -- | -- | -- | -- | -- | -- |
| | | | | | | | | | | | | |
| | | | | | | | | | | | | |
| Age (centered) | -0.11 | | | | | | | | | | | |
| | -0.11 | | | | | | | | | | | |
| | | | | | | | | | | | | |
| Female Gender | | 0.12 | | | | | | | | | | |
| | | -1.44 | | | | | | | | | | |
| | | | | | | | | | | | | |
| White Race | | | -3.50 | | | | | | | | | |
| | | | -1.92 | | | | | | | | | |
| | | | | | | | | | | | | |
| Conservative (Ref: Liberal) | | | | -0.83 | | | | | | | | |
| | | | | -1.62 | | | | | | | | |
| | | | | | | | | | | | | |
| Moderate (Ref: Liberal) | | | | -1.11 | | | | | | | | |
| | | | | -1.82 | | | | | | | | |
| | | | | | | | | | | | | |
| Suburb (Ref: Rural) | | | | | 1.45 | | | | | | | |
| | | | | | -1.64 | | | | | | | |
| | | | | | | | | | | | | |
| Urban (Ref: Rural) | | | | | 3.65 | | | | | | | |
| | | | | | -1.88 | | | | | | | |
| | | | | | | | | | | | | |
| Good Self Reported Health | | | | | | -0.83 | | | | | | |
| | | | | | | -1.72 | | | | | | |
| | | | | | | | | | | | | |
| High Trust in AI | | | | | | | 1.01 | | | | | |
| | | | | | | | -2.72 | | | | | |
| | | | | | | | | | | | | |
| Trust in CDC | | | | | | | | -0.14 | | | | |
| | | | | | | | | -0.56 | | | | |
| | | | | | | | | | | | | |
| Trust in NIH | | | | | | | | | -0.32 | | | |
| | | | | | | | | | -0.60 | | | |
| | | | | | | | | | | | | |
| Trust in PCP | | | | | | | | | | -0.10 | | |
| | | | | | | | | | | -0.84 | | |
| | | | | | | | | | | | | |
| Seen PCP within 2 years | | | | | | | | | | | -3.45* | |
| | | | | | | | | | | | -1.50 | |
| | | | | | | | | | | | | |
| Never talked to doctor about CRC screening | | | | | | | | | | | | -2.68 |
| | | | | | | | | | | | | -1.83 |
| | | | | | | | | | | | | |
| Pre-treatment Colonoscopy Intent | 0.87*** | 0.87*** | 0.88*** | 0.88*** | 0.87*** | 0.87*** | 0.87*** | 0.88*** | 0.88*** | 0.88*** | 0.87*** | 0.89*** |
| | -0.02 | -0.02 | -0.02 | -0.02 | -0.02 | -0.02 | -0.02 | -0.02 | -0.02 | -0.02 | -0.02 | -0.02 |
| | | | | | | | | | | | | |
| Chatbot:Age (centered) | -0.09 | | | | | | | | | | | |
| | -0.22 | | | | | | | | | | | |
| | | | | | | | | | | | | |
| Single AI Message:Age (centered) | -0.14 | | | | | | | | | | | |
| | -0.27 | | | | | | | | | | | |
| | | | | | | | | | | | | |
| Expert Material:Age (centered) | -0.03 | | | | | | | | | | | |
| | -0.17 | | | | | | | | | | | |
| | | | | | | | | | | | | |
| Chatbot:Female Gender | | 0.89 | | | | | | | | | | |
| | | -3.53 | | | | | | | | | | |
| | | | | | | | | | | | | |
| Single AI Message:Female Gender | | 2.01 | | | | | | | | | | |
| | | -3.07 | | | | | | | | | | |
| | | | | | | | | | | | | |
| Expert Material:Female Gender | | -0.19 | | | | | | | | | | |
| | | -2.43 | | | | | | | | | | |
| | | | | | | | | | | | | |
| Chatbot:White Race | | | -5.90 | | | | | | | | | |
| | | | -5.26 | | | | | | | | | |
| | | | | | | | | | | | | |
| Single AI Message:White Race | | | 1.06 | | | | | | | | | |
| | | | -3.72 | | | | | | | | | |
| | | | | | | | | | | | | |
| Expert Material:White Race | | | 6.64* | | | | | | | | | |
| | | | -3.11 | | | | | | | | | |
| | | | | | | | | | | | | |
| Chatbot:Conservative (Ref: Liberal) | | | | 9.45* | | | | | | | | |
| | | | | -3.98 | | | | | | | | |
| | | | | | | | | | | | | |
| Single AI Message:Conservative (Ref: Liberal) | | | | 0.99 | | | | | | | | |
| | | | | -3.37 | | | | | | | | |
| | | | | | | | | | | | | |
| Expert Material:Conservative (Ref: Liberal) | | | | 2.46 | | | | | | | | |
| | | | | -2.84 | | | | | | | | |
| | | | | | | | | | | | | |
| Chatbot:Moderate (Ref: Liberal) | | | | 9.83** | | | | | | | | |
| | | | | -3.96 | | | | | | | | |
| | | | | | | | | | | | | |
| Single AI Message:Moderate (Ref: Liberal) | | | | 3.50 | | | | | | | | |
| | | | | -3.75 | | | | | | | | |
| | | | | | | | | | | | | |
| Expert Material:Moderate (Ref: Liberal) | | | | 4.76 | | | | | | | | |
| | | | | -3.17 | | | | | | | | |
| | | | | | | | | | | | | |
| Chatbot:Suburb (Ref: Rural) | | | | | -5.82 | | | | | | | |
| | | | | | -3.99 | | | | | | | |
| | | | | | | | | | | | | |
| Single AI Message:Suburb (Ref: Rural) | | | | | 1.53 | | | | | | | |
| | | | | | -3.52 | | | | | | | |
| | | | | | | | | | | | | |
| Expert Material:Suburb (Ref: Rural) | | | | | 3.44 | | | | | | | |
| | | | | | -2.78 | | | | | | | |
| | | | | | | | | | | | | |
| Chatbot:Urban (Ref: Rural) | | | | | -7.78 | | | | | | | |
| | | | | | -4.42 | | | | | | | |
| | | | | | | | | | | | | |
| Single AI Message:Urban (Ref: Rural) | | | | | -3.80 | | | | | | | |
| | | | | | -4.63 | | | | | | | |
| | | | | | | | | | | | | |
| Expert Material:Urban (Ref: Rural) | | | | | -2.36 | | | | | | | |
| | | | | | -3.13 | | | | | | | |
| | | | | | | | | | | | | |
| Chatbot:Good Self Reported Health | | | | | | 4.84 | | | | | | |
| | | | | | | -3.78 | | | | | | |
| | | | | | | | | | | | | |
| Single AI Message:Good Self Reported Health | | | | | | 3.26 | | | | | | |
| | | | | | | -3.28 | | | | | | |
| | | | | | | | | | | | | |
| Expert Material:Good Self Reported Health | | | | | | -0.78 | | | | | | |
| | | | | | | -2.85 | | | | | | |
| | | | | | | | | | | | | |
| Chatbot:High Trust in AI | | | | | | | 4.00 | | | | | |
| | | | | | | | -4.29 | | | | | |
| | | | | | | | | | | | | |
| Single AI Message:High Trust in AI | | | | | | | 2.01 | | | | | |
| | | | | | | | -4.33 | | | | | |
| | | | | | | | | | | | | |
| Expert Material:High Trust in AI | | | | | | | 1.10 | | | | | |
| | | | | | | | -4.13 | | | | | |
| | | | | | | | | | | | | |
| Chatbot:Trust in CDC | | | | | | | | -3.75** | | | | |
| | | | | | | | | -1.44 | | | | |
| | | | | | | | | | | | | |
| Single AI Message:Trust in CDC | | | | | | | | 0.71 | | | | |
| | | | | | | | | -0.99 | | | | |
| | | | | | | | | | | | | |
| Expert Material:Trust in CDC | | | | | | | | 1.75 | | | | |
| | | | | | | | | -1.03 | | | | |
| | | | | | | | | | | | | |
| Chatbot:Trust in NIH | | | | | | | | | -2.21 | | | |
| | | | | | | | | | -1.40 | | | |
| | | | | | | | | | | | | |
| Single AI Message:Trust in NIH | | | | | | | | | 0.58 | | | |
| | | | | | | | | | -1.06 | | | |
| | | | | | | | | | | | | |
| Expert Material:Trust in NIH | | | | | | | | | 2.32* | | | |
| | | | | | | | | | -0.98 | | | |
| | | | | | | | | | | | | |
| Chatbot:Trust in PCP | | | | | | | | | | -3.55 | | |
| | | | | | | | | | | -1.94 | | |
| | | | | | | | | | | | | |
| Single AI Message:Trust in PCP | | | | | | | | | | 2.49 | | |
| | | | | | | | | | | -1.57 | | |
| | | | | | | | | | | | | |
| Expert Material:Trust in PCP | | | | | | | | | | 1.45 | | |
| | | | | | | | | | | -1.54 | | |
| | | | | | | | | | | | | |
| Chatbot:Seen PCP within 2 years | | | | | | | | | | | 7.81* | |
| | | | | | | | | | | | -3.68 | |
| | | | | | | | | | | | | |
| Single AI Message:Seen PCP within 2 years | | | | | | | | | | | 3.99 | |
| | | | | | | | | | | | -3.07 | |
| | | | | | | | | | | | | |
| Expert Material:Seen PCP within 2 years | | | | | | | | | | | 5.81* | |
| | | | | | | | | | | | -2.69 | |
| | | | | | | | | | | | | |
| Chatbot:Never talked to doctor about CRC screening | | | | | | | | | | | | 10.50** |
| | | | | | | | | | | | | -3.51 |
| | | | | | | | | | | | | |
| Single AI Message:Never talked to doctor about CRC screening | | | | | | | | | | | | 5.16 |
| | | | | | | | | | | | | -3.12 |
| | | | | | | | | | | | | |
| Expert Material:Never talked to doctor about CRC screening | | | | | | | | | | | | 6.48** |
| | | | | | | | | | | | | -2.55 |
| | | | | | | | | | | | | |
| Constant | 3.47*** | 3.30** | 6.18*** | 3.73*** | 1.73 | 3.98** | 3.28*** | 3.63 | 4.32 | 3.68 | 4.49*** | 4.68** |
| | -0.85 | -1.25 | -1.87 | -1.17 | -1.26 | -1.60 | -0.83 | -2.08 | -2.26 | -3.40 | -0.97 | -1.84 |
| | | | | | | | | | | | | |
| | | | | | | | | | | | | |
| Observations | 915.00 | 915.00 | 915.00 | 915.00 | 915.00 | 915.00 | 915.00 | 915.00 | 915.00 | 915.00 | 915.00 | 915.00 |
| R2 | 0.76 | 0.76 | 0.76 | 0.76 | 0.76 | 0.76 | 0.76 | 0.76 | 0.76 | 0.76 | 0.76 | 0.76 |
| Adjusted R2 | 0.76 | 0.76 | 0.76 | 0.76 | 0.76 | 0.76 | 0.76 | 0.76 | 0.76 | 0.76 | 0.76 | 0.76 |
| Residual Std. Error | 17.05 (df = 906) | 17.08 (df = 906) | 16.98 (df = 906) | 16.97 (df = 902) | 17.04 (df = 902) | 17.05 (df = 906) | 17.04 (df = 906) | 16.91 (df = 906) | 16.98 (df = 906) | 16.97 (df = 906) | 17.03 (df = 906) | 16.95 (df = 906) |
| F Statistic | 359.08*** (df = 8; 906) | 357.26*** (df = 8; 906) | 362.74*** (df = 8; 906) | 242.59*** (df = 12; 902) | 240.03*** (df = 12; 902) | 358.85*** (df = 8; 906) | 359.27*** (df = 8; 906) | 366.88*** (df = 8; 906) | 362.59*** (df = 8; 906) | 363.05*** (df = 8; 906) | 360.10*** (df = 8; 906) | 364.35*** (df = 8; 906) |
| | | | | | | | | | | | | |
| Note: | *p<0.1; **p<0.05; ***p<0.01 | | | | | | | | | | | |

**Table S8.** Stool Test Intent Moderator Analysis
OLS Regressions with Interaction. P-Values are BH corrected. SEs reported below coefficients.

| | Post-Treatment Stool Test Intent | | | | | | | | | | | |
|---|---|---|---|---|---|---|---|---|---|---|---|---|
| | | | | | | | | | | | | |
| Chatbot | 13.87*** | 13.80*** | 12.98* | 10.34*** | 20.06*** | 16.12*** | 13.61*** | 33.09*** | 28.29*** | 32.15*** | 12.72*** | 3.21 |
| | -2.03 | -3.24 | -6.02 | -3.03 | -4.40 | -4.14 | -2.17 | -6.38 | -5.97 | -8.54 | -2.58 | -3.72 |
| | | | | | | | | | | | | |
| Single AI Message | 12.87*** | 12.80*** | 16.12*** | 13.03*** | 8.95** | 12.46*** | 12.51*** | 8.03 | 8.55 | 3.96 | 11.07*** | 6.88 |
| | -1.85 | -2.82 | -4.42 | -2.54 | -3.27 | -3.30 | -2.13 | -4.89 | -5.32 | -8.03 | -2.16 | -3.59 |
| | | | | | | | | | | | | |
| Expert Material | 7.59*** | 6.92*** | 4.68 | 6.23*** | 6.09 | 10.94*** | 7.44*** | 1.26 | -1.59 | -4.84 | 4.92** | 2.82 |
| | -1.62 | -2.32 | -3.03 | -2.00 | -3.26 | -3.28 | -1.73 | -4.65 | -4.42 | -7.25 | -1.96 | -3.42 |
| | | | | | | | | | | | | |
| No Message | -- | -- | -- | -- | -- | -- | -- | -- | -- | -- | -- | -- |
| | | | | | | | | | | | | |
| | | | | | | | | | | | | |
| Age (centered) | -0.03 | | | | | | | | | | | |
| | -0.14 | | | | | | | | | | | |
| | | | | | | | | | | | | |
| Female Gender | | 2.00 | | | | | | | | | | |
| | | -1.80 | | | | | | | | | | |
| | | | | | | | | | | | | |
| White Race | | | -1.56 | | | | | | | | | |
| | | | -2.09 | | | | | | | | | |
| | | | | | | | | | | | | |
| Conservative (Ref: Liberal) | | | | -0.26 | | | | | | | | |
| | | | | -2.18 | | | | | | | | |
| | | | | | | | | | | | | |
| Moderate (Ref: Liberal) | | | | 2.38 | | | | | | | | |
| | | | | -2.40 | | | | | | | | |
| | | | | | | | | | | | | |
| Suburb (Ref: Rural) | | | | | -0.10 | | | | | | | |
| | | | | | -2.53 | | | | | | | |
| | | | | | | | | | | | | |
| Urban (Ref: Rural) | | | | | -1.48 | | | | | | | |
| | | | | | -2.66 | | | | | | | |
| | | | | | | | | | | | | |
| Good Self Reported Health | | | | | | 0.17 | | | | | | |
| | | | | | | -2.33 | | | | | | |
| | | | | | | | | | | | | |
| High Trust in AI | | | | | | | 2.75 | | | | | |
| | | | | | | | -2.32 | | | | | |
| | | | | | | | | | | | | |
| Trust in CDC | | | | | | | | 0.18 | | | | |
| | | | | | | | | -0.69 | | | | |
| | | | | | | | | | | | | |
| Trust in NIH | | | | | | | | | -0.05 | | | |
| | | | | | | | | | -0.71 | | | |
| | | | | | | | | | | | | |
| Trust in PCP | | | | | | | | | | 0.62 | | |
| | | | | | | | | | | -1.10 | | |
| | | | | | | | | | | | | |
| Seen PCP within 2 years | | | | | | | | | | | -1.78 | |
| | | | | | | | | | | | -2.10 | |
| | | | | | | | | | | | | |
| Never talked to doctor about CRC screening | | | | | | | | | | | | -7.28*** |
| | | | | | | | | | | | | -2.45 |
| | | | | | | | | | | | | |
| Pre-treatment Stool Test Intent | 0.82*** | 0.82*** | 0.82*** | 0.83*** | 0.82*** | 0.82*** | 0.82*** | 0.82*** | 0.82*** | 0.82*** | 0.82*** | 0.83*** |
| | -0.02 | -0.02 | -0.02 | -0.02 | -0.02 | -0.02 | -0.02 | -0.02 | -0.02 | -0.02 | -0.02 | -0.02 |
| | | | | | | | | | | | | |
| Chatbot:Age (centered) | 0.26 | | | | | | | | | | | |
| | -0.33 | | | | | | | | | | | |
| | | | | | | | | | | | | |
| Single AI Message:Age (centered) | -0.07 | | | | | | | | | | | |
| | -0.34 | | | | | | | | | | | |
| | | | | | | | | | | | | |
| Expert Material:Age (centered) | -0.06 | | | | | | | | | | | |
| | -0.22 | | | | | | | | | | | |
| | | | | | | | | | | | | |
| Chatbot:Female Gender | | -0.20 | | | | | | | | | | |
| | | -4.15 | | | | | | | | | | |
| | | | | | | | | | | | | |
| Single AI Message:Female Gender | | 0.02 | | | | | | | | | | |
| | | -3.71 | | | | | | | | | | |
| | | | | | | | | | | | | |
| Expert Material:Female Gender | | 0.99 | | | | | | | | | | |
| | | -3.20 | | | | | | | | | | |
| | | | | | | | | | | | | |
| Chatbot:White Race | | | 1.00 | | | | | | | | | |
| | | | -6.39 | | | | | | | | | |
| | | | | | | | | | | | | |
| Single AI Message:White Race | | | -3.75 | | | | | | | | | |
| | | | -4.84 | | | | | | | | | |
| | | | | | | | | | | | | |
| Expert Material:White Race | | | 3.49 | | | | | | | | | |
| | | | -3.56 | | | | | | | | | |
| | | | | | | | | | | | | |
| Chatbot:Conservative (Ref: Liberal) | | | | 8.83 | | | | | | | | |
| | | | | -4.72 | | | | | | | | |
| | | | | | | | | | | | | |
| Single AI Message:Conservative (Ref: Liberal) | | | | -1.20 | | | | | | | | |
| | | | | -4.23 | | | | | | | | |
| | | | | | | | | | | | | |
| Expert Material:Conservative (Ref: Liberal) | | | | 1.27 | | | | | | | | |
| | | | | -3.67 | | | | | | | | |
| | | | | | | | | | | | | |
| Chatbot:Moderate (Ref: Liberal) | | | | 3.45 | | | | | | | | |
| | | | | -4.93 | | | | | | | | |
| | | | | | | | | | | | | |
| Single AI Message:Moderate (Ref: Liberal) | | | | 1.80 | | | | | | | | |
| | | | | -4.91 | | | | | | | | |
| | | | | | | | | | | | | |
| Expert Material:Moderate (Ref: Liberal) | | | | 3.44 | | | | | | | | |
| | | | | -4.31 | | | | | | | | |
| | | | | | | | | | | | | |
| Chatbot:Suburb (Ref: Rural) | | | | | -8.94 | | | | | | | |
| | | | | | -5.23 | | | | | | | |
| | | | | | | | | | | | | |
| Single AI Message:Suburb (Ref: Rural) | | | | | 3.94 | | | | | | | |
| | | | | | -4.13 | | | | | | | |
| | | | | | | | | | | | | |
| Expert Material:Suburb (Ref: Rural) | | | | | 1.76 | | | | | | | |
| | | | | | -4.02 | | | | | | | |
| | | | | | | | | | | | | |
| Chatbot:Urban (Ref: Rural) | | | | | -6.18 | | | | | | | |
| | | | | | -5.75 | | | | | | | |
| | | | | | | | | | | | | |
| Single AI Message:Urban (Ref: Rural) | | | | | 7.68 | | | | | | | |
| | | | | | -5.40 | | | | | | | |
| | | | | | | | | | | | | |
| Expert Material:Urban (Ref: Rural) | | | | | 2.26 | | | | | | | |
| | | | | | -4.65 | | | | | | | |
| | | | | | | | | | | | | |
| Chatbot:Good Self Reported Health | | | | | | -3.36 | | | | | | |
| | | | | | | -4.73 | | | | | | |
| | | | | | | | | | | | | |
| Single AI Message:Good Self Reported Health | | | | | | 0.59 | | | | | | |
| | | | | | | -3.97 | | | | | | |
| | | | | | | | | | | | | |
| Expert Material:Good Self Reported Health | | | | | | -5.12 | | | | | | |
| | | | | | | -3.77 | | | | | | |
| | | | | | | | | | | | | |
| Chatbot:High Trust in AI | | | | | | | -0.09 | | | | | |
| | | | | | | | -5.65 | | | | | |
| | | | | | | | | | | | | |
| Single AI Message:High Trust in AI | | | | | | | 0.47 | | | | | |
| | | | | | | | -4.20 | | | | | |
| | | | | | | | | | | | | |
| Expert Material:High Trust in AI | | | | | | | -0.14 | | | | | |
| | | | | | | | -4.68 | | | | | |
| | | | | | | | | | | | | |
| Chatbot:Trust in CDC | | | | | | | | -5.46*** | | | | |
| | | | | | | | | -1.72 | | | | |
| | | | | | | | | | | | | |
| Single AI Message:Trust in CDC | | | | | | | | 1.39 | | | | |
| | | | | | | | | -1.37 | | | | |
| | | | | | | | | | | | | |
| Expert Material:Trust in CDC | | | | | | | | 1.81 | | | | |
| | | | | | | | | -1.33 | | | | |
| | | | | | | | | | | | | |
| Chatbot:Trust in NIH | | | | | | | | | -4.07** | | | |
| | | | | | | | | | -1.63 | | | |
| | | | | | | | | | | | | |
| Single AI Message:Trust in NIH | | | | | | | | | 1.23 | | | |
| | | | | | | | | | -1.45 | | | |
| | | | | | | | | | | | | |
| Expert Material:Trust in NIH | | | | | | | | | 2.65* | | | |
| | | | | | | | | | -1.27 | | | |
| | | | | | | | | | | | | |
| Chatbot:Trust in PCP | | | | | | | | | | -4.76* | | |
| | | | | | | | | | | -2.16 | | |
| | | | | | | | | | | | | |
| Single AI Message:Trust in PCP | | | | | | | | | | 2.25 | | |
| | | | | | | | | | | -2.07 | | |
| | | | | | | | | | | | | |
| Expert Material:Trust in PCP | | | | | | | | | | 3.21 | | |
| | | | | | | | | | | -1.87 | | |
| | | | | | | | | | | | | |
| Chatbot:Seen PCP within 2 years | | | | | | | | | | | 3.18 | |
| | | | | | | | | | | | -4.13 | |
| | | | | | | | | | | | | |
| Single AI Message:Seen PCP within 2 years | | | | | | | | | | | 6.49 | |
| | | | | | | | | | | | -4.15 | |
| | | | | | | | | | | | | |
| Expert Material:Seen PCP within 2 years | | | | | | | | | | | 8.12** | |
| | | | | | | | | | | | -3.50 | |
| | | | | | | | | | | | | |
| Chatbot:Never talked to doctor about CRC screening | | | | | | | | | | | | 16.12*** |
| | | | | | | | | | | | | -4.39 |
| | | | | | | | | | | | | |
| Single AI Message:Never talked to doctor about CRC screening | | | | | | | | | | | | 8.76* |
| | | | | | | | | | | | | -4.16 |
| | | | | | | | | | | | | |
| Expert Material:Never talked to doctor about CRC screening | | | | | | | | | | | | 6.99 |
| | | | | | | | | | | | | -3.86 |
| | | | | | | | | | | | | |
| Constant | 6.43*** | 5.18*** | 7.77*** | 5.74*** | 6.88*** | 6.27** | 6.15*** | 5.73* | 6.64** | 4.12 | 7.11*** | 11.24*** |
| | -1.13 | -1.37 | -1.91 | -1.48 | -2.21 | -2.18 | -1.20 | -2.55 | -2.59 | -4.14 | -1.24 | -2.51 |
| | | | | | | | | | | | | |
| | | | | | | | | | | | | |
| Observations | 915.00 | 915.00 | 915.00 | 915.00 | 915.00 | 915.00 | 915.00 | 915.00 | 915.00 | 915.00 | 915.00 | 915.00 |
| R2 | 0.62 | 0.62 | 0.62 | 0.63 | 0.63 | 0.63 | 0.62 | 0.63 | 0.63 | 0.63 | 0.63 | 0.63 |
| Adjusted R2 | 0.62 | 0.62 | 0.62 | 0.62 | 0.62 | 0.62 | 0.62 | 0.63 | 0.63 | 0.63 | 0.62 | 0.63 |
| Residual Std. Error | 21.51 (df = 906) | 21.50 (df = 906) | 21.50 (df = 906) | 21.43 (df = 902) | 21.47 (df = 902) | 21.48 (df = 906) | 21.50 (df = 906) | 21.25 (df = 906) | 21.33 (df = 906) | 21.34 (df = 906) | 21.44 (df = 906) | 21.35 (df = 906) |
| F Statistic | 188.42*** (df = 8; 906) | 188.71*** (df = 8; 906) | 188.66*** (df = 8; 906) | 127.51*** (df = 12; 902) | 126.68*** (df = 12; 902) | 189.21*** (df = 8; 906) | 188.72*** (df = 8; 906) | 195.82*** (df = 8; 906) | 193.61*** (df = 8; 906) | 193.10*** (df = 8; 906) | 190.27*** (df = 8; 906) | 192.99*** (df = 8; 906) |
| | | | | | | | | | | | | |
| Note: | *p<0.1; **p<0.05; ***p<0.01 | | | | | | | | | | | |

**Table S9.** Exploratory TOST analysis of the MI-style chatbot-minus-single-message contrast
Bounds are +/-0.20 times the outcome-specific residual SD. Equivalence requires both one-sided p values below .05 and the 90% CI entirely within the raw-unit bounds.

| Outcome | Residual SD | Margin | Chatbot-single | Standardized | 90% CI | 95% CI | p lower | p upper |
|---|---|---|---|---|---|---|---|---|
| Colonoscopy intent | 17.05 | +/-3.41 | -1.31 | -0.077 | -4.52 to 1.89 | -5.13 to 2.51 | 0.141 | 0.008 |
| Stool-test intent | 21.48 | +/-4.30 | 0.88 | 0.041 | -3.05 to 4.82 | -3.81 to 5.58 | 0.015 | 0.077 |

**Table S10.** Change-score models for immediate CRC screening intent

The dependent variable is post-intervention minus baseline intent on the 0-100 scale. Estimates are between-arm differences in mean change with HC2 robust standard errors.

| Outcome | Contrast | Estimate | Robust SE | 95% CI | p |
|---|---|---|---|---|---|
| Colonoscopy intent | MI-style AI chatbot vs Neutral-reading attention control | 4.18 | 1.72 | 0.805 to 7.555 | 0.015 |
| Colonoscopy intent | Single AI message vs Neutral-reading attention control | 5.77 | 1.48 | 2.871 to 8.662 | <.001 |
| Colonoscopy intent | Expert patient material vs Neutral-reading attention control | 5.91 | 1.23 | 3.492 to 8.327 | <.001 |
| Colonoscopy intent | MI-style AI chatbot vs Single AI message | -1.59 | 2.03 | -5.579 to 2.407 | 0.436 |
| Colonoscopy intent | MI-style AI chatbot vs Expert patient material | -1.73 | 1.87 | -5.391 to 1.932 | 0.354 |
| Colonoscopy intent | Single AI message vs Expert patient material | -0.14 | 1.64 | -3.368 to 3.082 | 0.931 |
| Stool-test intent | MI-style AI chatbot vs Neutral-reading attention control | 13.49 | 2.1 | 9.363 to 17.615 | <.001 |
| Stool-test intent | Single AI message vs Neutral-reading attention control | 12.64 | 1.89 | 8.940 to 16.345 | <.001 |
| Stool-test intent | Expert patient material vs Neutral-reading attention control | 8.43 | 1.65 | 5.198 to 11.661 | <.001 |
| Stool-test intent | MI-style AI chatbot vs Single AI message | 0.85 | 2.51 | -4.089 to 5.781 | 0.737 |
| Stool-test intent | MI-style AI chatbot vs Expert patient material | 5.06 | 2.34 | 0.468 to 9.651 | 0.031 |
| Stool-test intent | Single AI message vs Expert patient material | 4.21 | 2.15 | -0.002 to 8.429 | 0.05 |

**Table S11.** Odds ratios for crossing the screening intent threshold (≤50 → >50) by intervention arm, among participants with low baseline (≤50) intent.
SEs reported below coefficients

| | Colonoscopy (≤50 → >50) (Reference = Control) | Colonoscopy (≤50 → >50) (Reference = Expert Material) | Stool Test (≤50 → >50) (Reference = Control) | Stool Test (≤50 → >50) (Reference = Expert Material) | Colonoscopy (≤25 → >25) (Reference = Control) | Colonoscopy (≤25 → >25) (Reference = Expert Material) | Stool Test (≤25 → >25) (Reference = Control) | Stool Test (≤25 → >25) (Reference = Expert Material) | Colonoscopy (≤75 → >75) (Reference = Control) | Colonoscopy (≤75 → >75) (Reference = Expert Material) | Stool Test (≤75 → >75) (Reference = Control) | Stool Test (≤75 → >75) (Reference = Expert Material) |
|---|---|---|---|---|---|---|---|---|---|---|---|---|
| | | | | | | | | | | | | |
| No Message | -- | 0.43* | -- | 0.41** | -- | 0.35** | -- | 0.20*** | -- | 0.19** | -- | 0.50* |
| | | -0.51 | | -0.36 | | -0.41 | | -0.38 | | -0.78 | | -0.4 |
| | | | | | | | | | | | | |
| Chatbot | 4.36*** | 1.87* | 5.42*** | 2.20*** | 5.44*** | 1.93** | 10.52*** | 2.12*** | 4.67* | 0.9 | 5.07*** | 2.54*** |
| | -0.48 | -0.38 | -0.35 | -0.27 | -0.4 | -0.31 | -0.38 | -0.27 | -0.8 | -0.49 | -0.37 | -0.3 |
| | | | | | | | | | | | | |
| Single AI Message | 4.36*** | 1.86* | 5.28*** | 2.14*** | 4.32*** | 1.53 | 6.97*** | 1.41 | 6.01** | 1.15 | 3.93*** | 1.96** |
| | -0.48 | -0.37 | -0.35 | -0.27 | -0.41 | -0.32 | -0.39 | -0.27 | -0.78 | -0.46 | -0.38 | -0.31 |
| | | | | | | | | | | | | |
| Expert Material | 2.34* | | 2.47** | | 2.82** | | 4.95*** | | 5.21** | | 2.00* | |
| | -0.51 | | -0.36 | | -0.41 | | -0.38 | | -0.78 | | -0.4 | |
| | | | | | | | | | | | | |
| Constant | 0.03*** | 0.08*** | 0.07*** | 0.18*** | 0.06*** | 0.18*** | 0.08*** | 0.41*** | 0.01*** | 0.05*** | 0.05*** | 0.10*** |
| | -0.42 | -0.29 | -0.3 | -0.21 | -0.35 | -0.23 | -0.33 | -0.19 | -0.71 | -0.33 | -0.32 | -0.24 |
| | | | | | | | | | | | | |
| | | | | | | | | | | | | |
| Observations | 682 | 682 | 682 | 682 | 535 | 535 | 492 | 492 | 771 | 771 | 791 | 791 |
| Log Likelihood | -195.47 | -195.47 | -310.05 | -310.05 | -228.75 | -228.75 | -267.42 | -267.42 | -122.59 | -122.59 | -282.09 | -282.09 |
| Akaike Inf. Crit. | 398.94 | 398.94 | 628.11 | 628.11 | 465.5 | 465.5 | 542.85 | 542.85 | 253.18 | 253.18 | 572.19 | 572.19 |
| | | | | | | | | | | | | |
| Note: | *p<0.1; **p<0.05; ***p<0.01 | | | | | | | | | | | |

**Table S12.** Sensitivity of treatment effects to differential survey completion. Baseline-adjusted OLS estimates of post-intervention vaccination intention using baseline-observation-carried-forward imputation for participants missing the post-intervention outcome. The analysis includes forty-five otherwise eligible participants did not complete the post-intervention intention items (N = 960).

| Outcome | Contrast | Estimate | Robust SE | 95% CI | p value |
|---|---|---|---|---|---|
| Colonoscopy intent | MI-style AI chatbot vs no-message control | 4.23 | 1.57 | 1.16 to 7.30 | 0.007 |
| Stool-test intent | MI-style AI chatbot vs no-message control | 12.9 | 1.94 | 9.10 to 16.71 | <.001 |
| Colonoscopy intent | Single AI message vs no-message control | 5.48 | 1.36 | 2.81 to 8.15 | <.001 |
| Stool-test intent | Single AI message vs no-message control | 11.87 | 1.76 | 8.42 to 15.31 | <.001 |
| Colonoscopy intent | Expert patient material vs no-message control | 5.08 | 1.18 | 2.76 to 7.40 | <.001 |
| Stool-test intent | Expert patient material vs no-message control | 7.09 | 1.57 | 4.02 to 10.17 | <.001 |
| Colonoscopy intent | MI-style AI chatbot vs expert patient material | -0.85 | 1.7 | -4.18 to 2.48 | 0.617 |
| Stool-test intent | MI-style AI chatbot vs expert patient material | 5.81 | 2.14 | 1.61 to 10.01 | 0.007 |
| Colonoscopy intent | Single AI message vs expert patient material | 0.39 | 1.52 | -2.59 to 3.37 | 0.795 |
| Stool-test intent | Single AI message vs expert patient material | 4.77 | 1.99 | 0.87 to 8.67 | 0.016 |
| Colonoscopy intent | MI-style AI chatbot vs single AI message | -1.24 | 1.83 | -4.84 to 2.35 | 0.497 |
| Stool-test intent | MI-style AI chatbot vs single AI message | 1.04 | 2.27 | -3.42 to 5.49 | 0.648 |

**Table S13.** Regression Results Comparing Chatbot and Public Health Material Effectiveness SEs reported below coefficients.

| | Comfort discussing CRC screening with PCP (Reference = No message) | Comfort discussing CRC screening with PCP (Reference = Expert Material) | How Empathetic was the Intervention? (out of 5) (Reference = Expert Material) | Message Effectiveness (out of 20) (Reference = Expert Material) |
|---|---|---|---|---|
| | | | | |
| No Message | -- | -1.35 | | |
| | | -2.72 | | |
| | | | | |
| Chatbot | 8.12*** | 6.77*** | 0.85*** | 0.33 |
| | -2.60 | -2.52 | -0.10 | -0.28 |
| | | | | |
| Single AI Message | 6.94*** | 5.58** | 0.55*** | 0.70** |
| | -2.63 | -2.55 | -0.10 | -0.29 |
| | | | | |
| Expert Material | 1.35 | -- | -- | -- |
| | -2.72 | | | |
| | | | | |
| Constant | 70.04*** | 71.39*** | 2.54*** | 14.78*** |
| | -1.98 | -1.87 | -0.07 | -0.20 |
| | | | | |
| | | | | |
| Observations | 915.00 | 915.00 | 668.00 | 668.00 |
| R2 | 0.02 | 0.02 | 0.10 | 0.01 |
| Adjusted R2 | 0.01 | 0.01 | 0.10 | 0.01 |
| | | | | |
| Note: | | | | *p<0.1; **p<0.05; ***p<0.01 |

**Table S14.** Surplus Time Spent by Arm (seconds)

| | **No Message N = 241** | **Chatbot N = 211** | **Single AI Message N = 206** | **Expert Patient Material N = 221** | **p-value[1]** |
|---|---|---|---|---|---|
| Surplus Time Spent Over 3 minute floor, seconds (mean (SD)) | 44.32 (105.79) | 213.42 (260.46) | 33.41 (69.61) | 37.25 (78.46) | <0.001 |

[1]Kruskal-Wallis rank sum test

**Table S15.** Single AI Message Persuasion Strategy Usage

| Persuasion Strategy | Mean Usage (Std) (0: None, 3: High) | Theme |
|---|---|---|
| Use of Direct Address | 3.0 (0.0) | Rhetorical |
| Emphasize Empowerment and Personal Agency | 2.98 (0.15) | Identity Appeals |
| Appeal to Age Appropriateness | 2.83 (0.37) | Identity Appeals |
| Appeal to Independence/Self-Sufficiency/Personal Responsibility | 2.74 (0.44) | Identity Appeals |
| Connect to Long-Term Health | 2.7 (0.46) | Emotional Appeals |
| Appeal to Science and Evidence-Based Action | 2.64 (0.48) | Identity Appeals |
| Appeal to Family Responsibility | 2.47 (0.52) | Social Norms |
| Provide Clear Next Steps | 2.39 (0.49) | Practical/Process |
| Highlight Reduced Cancer Risk | 2.33 (0.47) | Facts and Information |
| Appeal to Self-Identity Appeals | 2.24 (0.43) | Identity Appeals |
| Appeal to Common Sense/Practicality | 2.14 (0.4) | Facts and Information |
| Normalize Routine Screening | 2.05 (0.21) | Practical/Process |
| Focus on Non-Invasive Choices | 2.03 (0.17) | Practical/Process |
| Reference Doctor Recommendations | 2.01 (0.17) | Facts and Information |
| Appeal to Self-Respect | 2.0 (0.18) | Identity Appeals |
| Emphasize Ease of Stool Tests | 1.96 (0.25) | Practical/Process |
| Counter Screening Misconceptions | 1.94 (0.32) | Practical/Process |
| Highlight Screening Simplicity | 1.94 (0.29) | Practical/Process |
| Highlight Endorsements from Trusted Organizations | 1.92 (0.35) | Facts and Information |
| Reassure on Procedure Safety | 1.89 (0.34) | Facts and Information |
| Emphasize Peace of Mind | 1.87 (0.44) | Emotional Appeals |
| Highlight Role Modeling for Others | 1.87 (0.59) | Social Norms |
| Appeal to Community Norms or Peer Behavior | 1.85 (0.53) | Social Norms |
| Address Common Fears | 1.82 (0.4) | Emotional Appeals |
| Appeal to Gender Roles | 1.73 (0.81) | Identity Appeals |
| Appeal to Milestones/Life Events | 1.5 (0.57) | Identity Appeals |
| Highlight One-Time or Infrequent Need | 1.37 (0.88) | Practical/Process |
| Reassure About Minimal Discomfort | 1.33 (0.55) | Practical/Process |
| Appeal to Social Justice or Health Equity | 1.11 (1.17) | Identity Appeals |
| Cite Screening Statistics | 1.05 (0.77) | Facts and Information |
| Appeal to Rural or Urban Identity/Challenges | 1.03 (1.01) | Identity Appeals |
| Appeal to Education/Intelligence | 1.0 (0.94) | Identity Appeals |
| Present Easy Scheduling Options | 0.95 (0.67) | Practical/Process |
| Share Statistics on Success | 0.86 (0.63) | Facts and Information |
| Appeal to Active Lifestyle | 0.66 (0.74) | Identity Appeals |
| Appeal to Enjoyment of Retirement/Leisure | 0.59 (0.72) | Identity Appeals |
| Explain Cost and Insurance Coverage Details | 0.5 (0.82) | Facts and Information |
| Spotlight Local Screening Access | 0.5 (0.72) | Practical/Process |
| Emphasize Pros of Flexible Sigmoidoscopy | 0.49 (0.51) | Practical/Process |
| Appeal to Career Impact | 0.48 (0.66) | Identity Appeals |
| Promote Free or Low-Cost Options | 0.43 (0.72) | Practical/Process |
| Counter Embarrassment Stigma | 0.36 (0.58) | Emotional Appeals |
| Reassure Privacy and Dignity | 0.36 (0.57) | Practical/Process |
| Use of Analogies | 0.21 (0.55) | Facts and Information |
| Appeal to Patriotism/National Identity Appeals | 0.14 (0.47) | Identity Appeals |
| Appeal to Use of Earned Benefits | 0.12 (0.38) | Identity Appeals |
| Mention Quick Results Timeline | 0.02 (0.15) | Facts and Information |
| Focus on Quick Recovery | 0.02 (0.15) | Practical/Process |
| Share Testimonials from Survivors | 0.0 (0.0) | Emotional Appeals |